\documentclass[desactivate]{aa}  

\usepackage{natbib}
\bibpunct{(}{)}{;}{a}{}{,}

\usepackage{graphicx}
\usepackage{revsymb}	
\usepackage{amsmath}
\usepackage{txfonts}
\usepackage{amssymb}
\usepackage{geometry} 
\usepackage[parfill]{parskip} 
\usepackage{slantsc}
\usepackage{array}
\usepackage{amsfonts}
	
\usepackage{epstopdf}	
\usepackage{epsfig}	

\usepackage{subfigure} 
\usepackage{subcaption}
\usepackage{url}

\usepackage[colorlinks=true,linkcolor=black, urlcolor=black, citecolor=black]{hyperref}

\usepackage{color}

\usepackage{xfrac}

\usepackage{sidecap}
\usepackage[font=small, labelfont=bf]{caption}
\usepackage{multirow}
\usepackage{booktabs}
\usepackage{threeparttable}

\newcommand{\figsep}{\par\vspace{1mm}}

\begin{document}
   \title{Detailed study of HD~219134: A benchmark K dwarf with a compact planetary system}
\author{J. B\'{e}trisey\inst{1,2} \and C. Pezzotti\inst{3} \and G. Buldgen\inst{4} \and A.~A. Vidotto\inst{5} \and C.~P. Folsom\inst{6} \and M. De~Becker\inst{4} \and M. Gillon\inst{7}}
\institute{
Institut für Astrophysik, Universität Zürich, Winterthurerstrasse 190, 8057 Zürich, Switzerland\\email: 	\texttt{jerome.betrisey@uzh.ch}
\and Department of Physics and Astronomy, Uppsala University, Box 516, SE-751 20 Uppsala, Sweden
\and INAF-IAPS, Via del Fosso del Cavaliere 100, Roma, Italy
\and STAR Institute, University of Liège, Allée du 6 Août 19C, 4000 Liège, Belgium
\and  Leiden Observatory, Leiden University, P.O. Box 9513, NL-2300 RA Leiden, The Netherlands
\and Tartu Observatory, University of Tartu, Observatooriumi 1, Tõravere, 61602 Tartumaa, Estonia
\and Astrobiology Research Unit, University of Liège, Allée du 6 Août 19C, 4000 Liège, Belgium}
\date{\today}

\abstract
{Space-based missions such as CoRoT, \textit{Kepler}, and TESS have revolutionised asteroseismology and strengthened its synergies with exoplanet science. With PLATO and ARIEL on the horizon, these synergies will play an even more central role.}
{We revisit the characterisation of the nearby K dwarf HD~219134 using advanced seismic diagnostics. Our goals are to investigate the reported tension between asteroseismic and interferometric radii, update the fundamental parameters of the star, refine the properties of its planetary system, and investigate the coupled rotational, magnetic, and star–planet evolution of the system, including the atmospheric mass‑loss histories of its planets.}
{The stellar characterisation is based on detailed asteroseismic forward modelling complemented by seismic inversions to quantify systematic uncertainties associated with non‑seismic constraints, assess the compatibility of radius measurements, and evaluate the impact of surface modelling on seismic inferences. The revised stellar properties were used as inputs for rotational evolution investigations, stellar wind modelling, and high‑energy irradiation histories.}
{Seismic inversions rule out the most recent interferometric radius determination, whereas asteroseismic and spectroscopic results are compatible with the previous interferometric measurement from 2019. We derive revised stellar mass, radius, and age of $0.779 \pm 0.015_\mathrm{(stat)}\pm 0.006_\mathrm{(sys)}\, M_\odot$, $0.753 \pm 0.005_\mathrm{(stat)}\pm 0.002_\mathrm{(sys)} \, R_\odot$, and $8.92 \pm 0.89_\mathrm{(stat)}\pm 0.19_\mathrm{(sys)}$ Gyr, respectively. The quoted systematic uncertainties reflect contributions from the non-seismic observational constraints only. Literature values for combined uncertainties arising from physical inputs and surface-effect prescriptions are $0.007M_\odot$, $0.002R_\odot$, and 0.81 Gyr, respectively. Using stellar wind models and a dedicated calibration of the magnetic braking law for low-mass stars, we reconstructed the rotational and angular‑momentum evolution of the host star and find that standard magnetic braking prescriptions remain broadly consistent with the observed present‑day rotation and angular‑momentum loss. Models invoking weakened magnetic braking provide a closer match but challenge commonly adopted scaling relations for the wind-driven angular momentum loss. The reconstructed X‑ray activity history implies early atmospheric stripping of the innermost planets, while the more massive outer planets are expected to have retained at least part of their primordial envelopes. In particular, HD~219134d appears to have undergone only moderate atmospheric erosion over the system lifetime.}
{The available evidence suggests that the apparent radius discrepancy for HD~219134 does not reflect a genuine astrophysical problem but is more likely the result of underestimated uncertainties in previous interferometric analyses. We therefore advocate adopting the asteroseismic radius, which demonstrates remarkable robustness against variations in modelling assumptions and optimisation strategies.}


\maketitle

\defcitealias{Ligi2019}{L19}
\defcitealias{Rosenthal2021}{R21}
\defcitealias{Li2025}{L25}
\defcitealias{Gillon2017_HD219134}{G17}

\section{Introduction}
Asteroseismology, the study of global stellar oscillations, has transformed our ability to characterise solar‑type stars \citep[see e.g. the reviews by][]{Chaplin&Miglio2013,Garcia&Ballot2019,Aerts2021}. By probing the oscillation modes that propagate throughout their interiors, asteroseismic analyses yield stringent constraints on stellar structure and fundamental parameters, including radius, mass, and age. Over the past two decades, the rapid development of this field has had profound implications for the study of planetary systems, whose detection and characterisation greatly benefit from accurate knowledge of their host stars \citep[see e.g.][]{Huber2018}. Additionally, stars and planets interact through multiple mechanisms during their evolution \citep[see e.g.][]{Lanza2015,Vidotto2020}, which are closely connected to the structural and rotational properties of the host star \citep[e.g.][]{Privitera2016_spi1,Privitera2016_RG,Privitera2016_spi2,Rao2018,Barker2020,Pezzotti2021_K444,Ahuir2021,Pezzotti2022,Pezzotti2025,Betrisey2022,Fellay2023}. A detailed understanding of the present state and evolutionary history of the host star is therefore essential for accurately characterising its planetary system and reconstructing its evolution.

In this context, K‑dwarf star-planet systems are particularly relevant. Compared to solar analogues, K dwarfs have smaller radii, enhancing transit detectability, and their typically lower masses improve radial‑velocity (RV) sensitivity. K dwarfs are also compelling targets in the search for habitable worlds: their long main‑sequence lifetimes provide extended periods of radiative stability that may favour planetary habitability and biological development. Their magnetic activity operates on a far smaller scale than that of M dwarfs, reducing both observational biases and the direct impact of high‑energy emission on planetary atmospheres. Moreover, their habitable zones lie closer to the host star than in G dwarfs, which enhances the detectability of temperate Earth‑sized planets with both transit and RV methods \citep[e.g.][]{LilloBox2022}. Understanding how the magnetic activity and the underlying dynamo processes of these stars evolve across their lifetimes is therefore crucial, given the direct consequences for star-planet interactions and atmospheric escape \citep{Wright2011,Kubyshkina2018,Kubyshkina2021,Pezzotti2026}. Obtaining precise ages for K dwarfs, however, remains challenging. Their lower luminosities and effective temperatures lead to reduced oscillation amplitudes, rendering oscillation-mode detections difficult with current instrumentation. The field is, however, approaching a turning point. While only a handful of seismic detections in early K stars have been achieved with past space-based missions \citep[e.g.][]{Barclay2013,Hon2024,Lund2025}, the landscape will considerably benefit from the capabilities of the forthcoming PLAnetary Transits and Oscillations of stars mission \citep[PLATO;][]{Rauer2025}. Additionally, the development of ground‑based extreme precision RV instrumentation has extended the detection boundary into the mid‑K range \citep[see e.g. discussion in][]{Li2025}. Moreover, to date, roughly a dozen K‑dwarf stars have reported seismic detections \citep[see e.g.][for a non-exhaustive list of references]{Carrier&Bourban2003,Kjeldsen2005,Carrier&Eggenberger2006,Campante2024,Lundkvist2024,Hon2024,Li2025}.

In this work, we focus on HD~219134, also known as HR~8832 or TOI‑1469. This star-planet system is particularly compelling for several reasons. First, at a distance of only 6.54\,pc \citep{Gaia2021}, HD~219134 is a bright ($V=5.57$) K3V host star with extensive and diverse observational coverage. Consequently, the system has been the subject of numerous studies \citep[e.g.][hereafter G17 and L25]{Gillon2017_HD219134,Li2025}, yielding an exceptionally rich dataset. HD~219134 has been characterised through a wide range of observational techniques, including measurements of magnetic activity, stellar winds, high‑energy irradiation, and rotation \citep{Marsden2014,Motalebi2015,Folsom2018}. Interferometric \citep[][Elliott et al., in prep.; hereafter E26. The stellar radius reported in E26 was previously communicated privately and adopted by \citetalias{Li2025}, whose value we use in this study]{Boyajian2012,Ligi2019} and spectroscopic \citep[][hereafter R21]{Valenti&Fischer2005,Rosenthal2021} observations further contribute to the detailed picture of this nearby K dwarf. The star also exhibits an 11.7 yr chromospheric activity cycle \citep{Johnson2016}, comparable in length to the solar cycle, making it a valuable analogue for studies of magnetic variability.

Moreover, HD~219134 is an old main‑sequence star \citep{Takeda2007,Gillon2017_HD219134,Li2025} and thus plays an important role in calibrating activity-rotation-age relations in the poorly constrained regime of older, slowly rotating stars. Its planetary system adds further interest: HD~219134 is one of the nearest and brightest stars known to host a compact, multiple transiting planet system \citep{Motalebi2015,Vogt2015,Gillon2017_HD219134,Rosenthal2021,Seager2021,Kokori2023}. The system is believed to host at least six planets \citep[see][and references therein]{Dietrich2022}, although only four have been robustly confirmed to date. Among these are the two transiting super-Earths, HD~219134b and HD~219134c, whose primary and secondary atmospheres are expected to have undergone substantial escape driven by intense high-energy irradiation \citep[e.g.][]{Kubyshkina2018}, which may have facilitated the development of refractory-rich atmospheres \citep{Vidotto2018}. In contrast, outermost planets may have preserved part of their primordial atmospheres.

Recent progress in the characterisation of HD~219134 has been driven by the detection of 25 acoustic oscillation modes with the Keck Planet Finder \citepalias{Li2025}, making HD~219134 one of the coolest stars with seismic detections to date. However, their analysis, which already had access to the interferometric radius from E26, reveals a significant tension between the asteroseismic radius and this new measurement. According to their investigation, this discrepancy cannot be attributed to known interferometric systematics, modelling assumptions (e.g. atmospheric boundary conditions or mixing-length theory assumptions), or commonly invoked non‑standard physical processes (e.g. magnetic activity or tidal heating). In this work, we revisit the characterisation of HD~219134 using complementary seismic analyses and more advanced modelling approaches, including seismic inversions. We discuss propagating the revised stellar parameters to update the planetary properties and examine the atmospheric mass loss of the system’s close‑in planets. The structure of the paper is as follows. In Sect.~\ref{sec:asteroseimic_modelling}, we describe the observational constraints and outline our seismic modelling strategy. Then, we discuss our assessment of the consistency between the interferometric and asteroseismic radii and present a new optimal stellar model for HD~219134. Section~\ref{sec:revision_planetary_parameters} presents the updated planetary parameters based on our revised stellar properties. In Sect.~\ref{sec:stellar_rotation_and_planet_mass_loss}, we revise the properties of the stellar wind and wind-driven mass loss of HD~219134. We investigate its rotational and X-ray luminosity evolution and discuss testing the impact of photoevaporation on the planetary atmospheres. Finally, we summarise our conclusions in Sect.~\ref{sec:conclusions}.

\section{Asteroseismic modelling}
\label{sec:asteroseimic_modelling}

\subsection{Observational constraints and modelling strategy}
\label{sec:obervational_constraints_and_modelling_strategy}
As observational constraints, we adopted the acoustic oscillation frequencies reported by \citetalias{Li2025}. In addition, HD~219134 benefits from high-quality spectroscopic and interferometric measurements \citepalias[see Table~3 in][]{Li2025}. \citetalias{Li2025} also derived the absolute luminosity using the distance from \citet{Bailer-Jones2021}. However, as noted by \citetalias{Li2025}, the two most recent interferometric determinations exhibit a significant inconsistency. Specifically, \citet[][hereafter L19]{Ligi2019} measured a radius of $0.728\pm 0.015~R_\odot$, whereas E26 report $0.783\pm 0.005~R_\odot$, a value incompatible with the former. A similar tension arises for the interferometric effective temperature: E26 obtained $4678\pm 45$~K, while \citetalias{Ligi2019} reported $4854\pm 66$~K, which agrees better with the spectroscopic estimate of $4817\pm 62$~K by \citetalias{Rosenthal2021}. Literature values of the non-seismic constraints (which are commonly referred to as classical constraints in the asteroseismic literature) are listed in Table~\ref{tab:set_constraints}.

\begin{table}[t!]
\centering
\caption{Classical constraints for HD~219134 from the literature.}
\resizebox{\linewidth}{!}{
\begin{tabular}{lc}
\hline \hline 
Classical constraint &  \\ 
\hline 
\textit{Interferometric radius} &  \\
\citet{Ligi2019} & $R_{L19}=0.728\pm 0.015$ $R_\odot$ \\
Elliott et al., in prep. & $R_{E26}=0.783 \pm 0.005$ $R_\odot$ \\
Elliott et al., in prep., degraded & $R_{E26}^{deg.}=0.783 \pm 0.013$ $R_\odot$ \\
\noalign{\vskip 0.3ex}
\hline
\textit{Absolute luminosity} &  \\
\citet{Li2025} & $L_{L25}=0.265\pm 0.011$ $L_\odot$ \\
\hline
\textit{Effective temperature} &  \\
\citet{Ligi2019} & $T_{\mathrm{eff}, L19} = 4854\pm 66$ K \\
\citet{Rosenthal2021} & $T_{\mathrm{eff}, R21} = 4817\pm 62$ K \\
Elliott et al., in prep. & $T_{\mathrm{eff}, E26} = 4678\pm 45$ K \\
\hline
\textit{Metallicity} &  \\
\citet{Rosenthal2021} & $\mathrm{[Fe/H]}_{R21} = 0.083\pm 0.058$ dex\\
\hline 
\end{tabular}}
\label{tab:set_constraints}
{\par\small\justify\textbf{Notes.} For the E26 data, we used the values provided in \citetalias{Li2025}. The precision of the degraded interferometric radius from E26 corresponds to the precision obtained for the 16 Cygni binary system, including the best characterised asteroseismic targets \citep[see e.g.][and references therein]{Buldgen2022b}. \par}
\end{table}

To infer the stellar properties, we employed two complementary approaches: forward modelling and seismic inversions \citep[see e.g. reviews by][]{Chaplin&Miglio2013,Buldgen2022c}. Forward modelling was carried out using the Asteroseismic Inference on a Massive Scale software \citep[AIMS;][]{Rendle2019} with the Spelaion grid \citep[][]{Betrisey2023_AMS_surf,Betrisey2026_catalog}, following the Forward and Inverse COmbination procedure \citep[FICO;][]{Betrisey2022,Betrisey2023_AMS_surf,Betrisey2024_AMS_quality,Betrisey2026_catalog}. Due to the sparse seismic data and missing modes, we could not compute enough frequency-separation ratios to apply the full FICO pipeline. Consequently, our inference relies on the first two steps of FICO: (i) direct fitting of individual frequencies combined with classical constraints, and (ii) seismic inversions. The inversions were performed using the automated version of InversionKit \citep[shortened as IK in the figures;][]{Reese&Zharkov2016}. We refer to \citet{Buldgen2022c} for a comprehensive review of seismic inversions \citep[see e.g.][for a non-exhaustive selection of references]{DiMauro2004_theo,Buldgen2015a,Buldgen2015b,Buldgen2022b,Buldgen2025_Lithium,Kosovichev&Kitiashvili2020,Salmon2021,Betrisey&Buldgen2022,Betrisey2023_rot,Betrisey2023_AMS_surf,Betrisey2024_AMS_quality,Betrisey2025_MA_Inv}. 

The Spelaion evolutionary tracks were generated under what we hereafter refer to as standard physics. Stellar models were computed with the Liège stellar evolution code \citep[CLES;][]{Scuflaire2008b}. For each evolutionary model, adiabatic oscillation frequencies were subsequently obtained using the Liège Oscillation Code \citep[LOSC;][]{Scuflaire2008a}. The chemical mixture adopted corresponds to AGSS09 abundances \citep{Asplund2009}. OPAL opacity tables \citep{Iglesias&Rogers1996} were employed, complemented at low temperatures by the data of \citet{Ferguson2005} and electron conductivity was treated following \citet{Potekhin1999}. Thermodynamic properties were described using the FreeEOS equation of state \citep{Irwin2012}, noting that OPAL opacity tables are themselves based on the OPAL2005 EOS \citep{Rogers&Nayfonov2002}. Microscopic diffusion was implemented according to the prescription of \citet{Thoul1994}, with screening coefficients from \citet{Paquette1986}. Nuclear reaction rates follow the determinations of \citet{Adelberger2011}. Mixing-length formalism from \citet{Cox&Giuli1968} was used to model convection, adopting a constant solar-calibrated mixing-length parameter of $\alpha_{\mathrm{MLT}} = 2.05$. For atmospheric modelling, we employed the $T(\tau)$ relation from model C of \citet{Vernazza1981}.

\subsection{Compatibility of interferometric and asteroseismic radii}
\label{sec:compatibility_radii}
In this section, we assess the agreement between interferometric and asteroseismic determinations of HD~219134 radius using forward-modelling and seismic inversion techniques. To achieve this, we complement the seismic data with ten distinct sets of classical constraints derived from recent measurements by \citetalias{Ligi2019}, \citetalias{Rosenthal2021}, \citetalias{Li2025}, and E26 (see Table~\ref{tab:set_constraints}). Given the exceptionally high precision of the E26 radius measurement, we constructed two additional sets where the interferometric radius is artificially degraded to match the precision obtained for the 16 Cygni binary system, among the best-characterised asteroseismic targets \citep[see e.g.][and references therein]{Buldgen2022b}. The complete list of constraint sets is provided in Table~\ref{tab:chi2_radius}.

\begin{figure}[t!]
\centering
\includegraphics[width=0.95\linewidth]{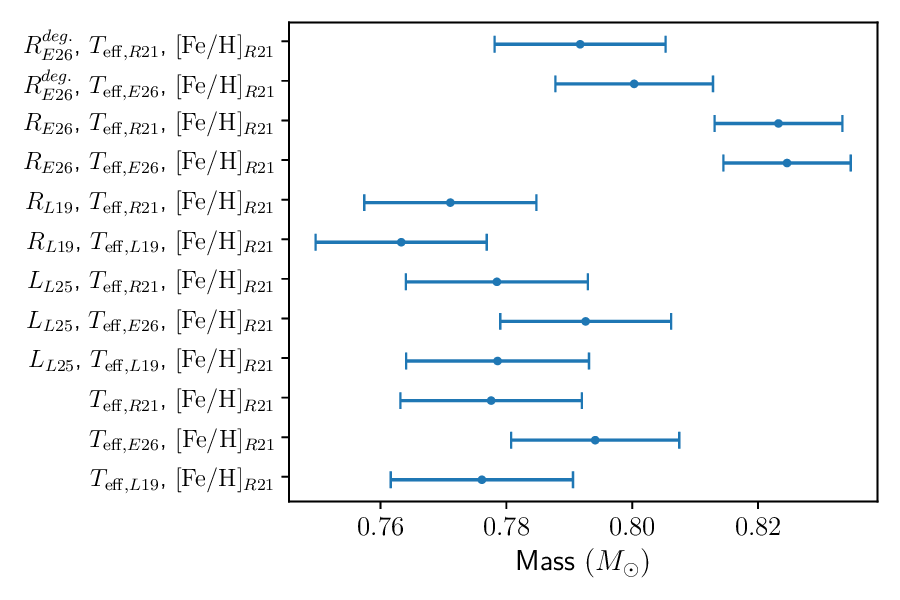}\figsep
\includegraphics[width=0.95\linewidth]{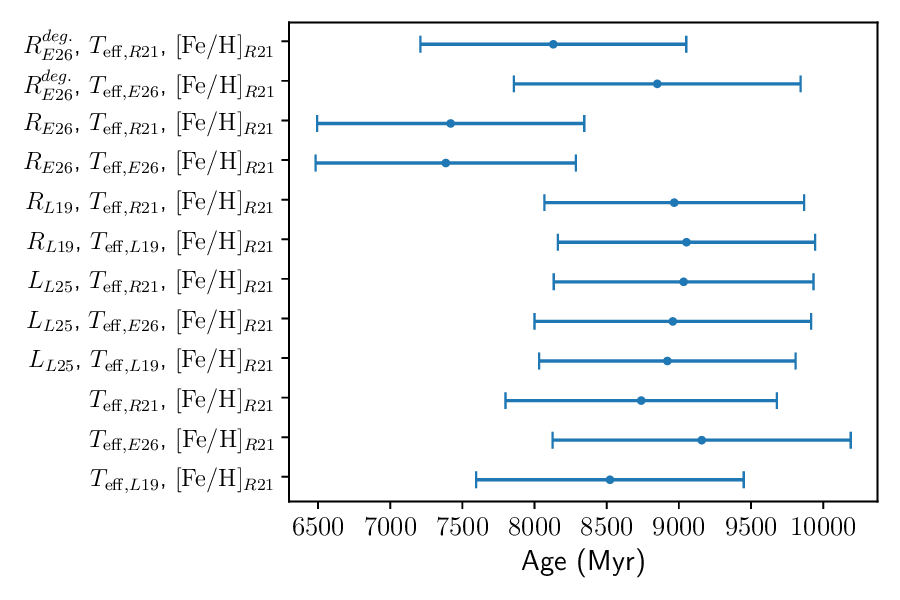}\figsep
\includegraphics[width=0.95\linewidth]{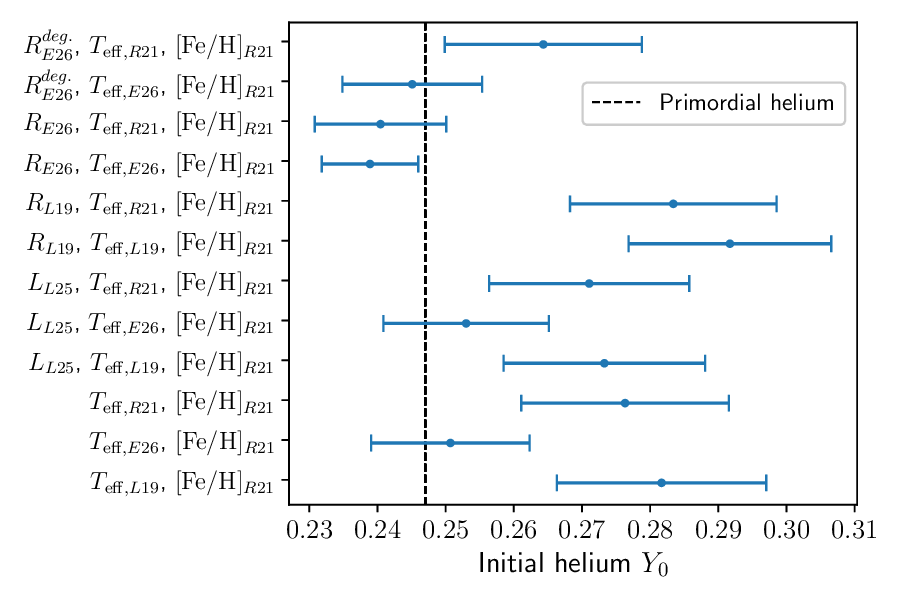}
\caption{\textit{From top to bottom:} Results from asteroseismic forward modelling, showing the inferred stellar mass, age, and initial helium mass fraction $Y_0$ for the 12 different sets of classical constraints. The dashed black line shows the primordial helium abundance $Y_p\simeq 0.247$ from \citet{Pitrou2018}.}
\label{fig:results_forward_modelling_MAY}
\end{figure}

For forward-modelling investigations, all twelve constraint sets were fitted along with the seismic data using the AIMS software, adopting the surface effect prescription of \citet{Ball&Gizon2014}. Figure~\ref{fig:results_forward_modelling_MAY} presents the inferred stellar properties, mass, age, and initial helium fraction $Y_0$. Models incorporating E26 data systematically yield lower initial helium fractions compared to those based on \citetalias{Ligi2019} or sets without interferometric constraints. In fact, the two models using the E26 interferometric radius fall below the primordial helium abundance, signalling a potential inconsistency between E26 measurements and the standard physics implemented in our stellar models. This discrepancy arises because the high precision of the E26 radius strongly influences the cost function, prompting AIMS to prioritise its matching. Unable to reconcile this constraint within the adopted physics, the minimisation compensates by driving $Y_0$ to non-physical values, which in turn also significantly biases the inferred mass and age. When the interferometric radius is degraded, reducing its weight in the cost function, the solutions converge in the direction of those obtained from the other constraint sets. We note that invoking non-standard physics is unlikely to resolve the radius discrepancy, as such mechanisms are not expected to induce variations of the required magnitude for similar stars \citep[e.g.][]{Buldgen2019f,Betrisey2022,Khan2026}, nor for HD~219134 specifically \citepalias{Li2025}.

\begin{figure}[t!]
\centering
\includegraphics[scale=0.6]{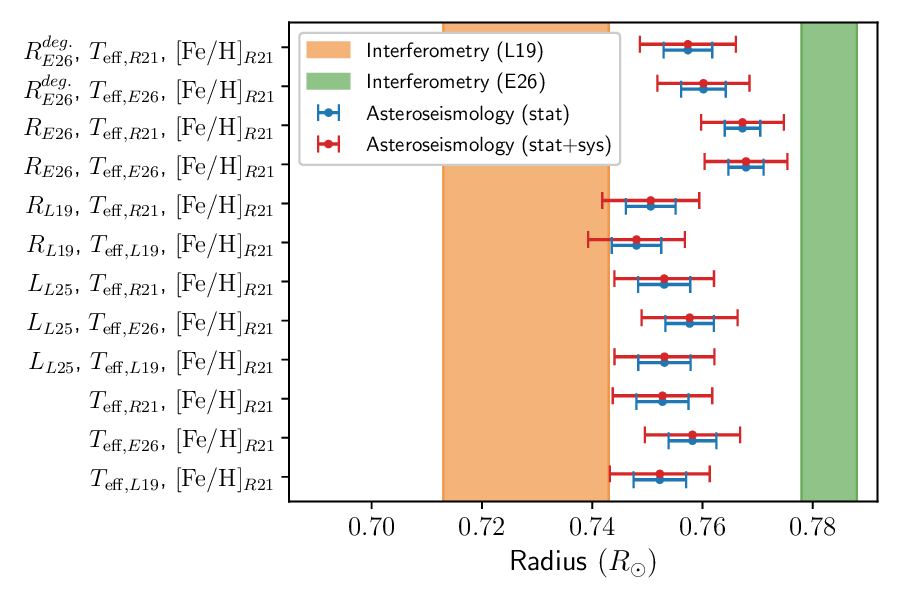}
\caption{Comparison between interferometric radii from \citetalias{Ligi2019} (orange) and E26 (green), and asteroseismology, including systematic uncertainties (red) or not (blue).}
\label{fig:comp_radius_interfero_astero}
\end{figure}

Figure~\ref{fig:comp_radius_interfero_astero} and Table~\ref{tab:chi2_radius} summarise the compatibility analysis of radius estimates from forward modelling, considering both statistical uncertainties alone (stat) and combined statistical plus systematic uncertainties (stat+sys). To quantify this, we defined a compatibility indicator $\chi^2$, where values below 2 indicate agreement (see Appendix~\ref{app:compatibility_indicator}).

When only statistical uncertainties are considered, none of the models are compatible with the E26 radius, and only one model agrees with the \citetalias{Ligi2019} measurement, unsurprisingly, the model that includes both the \citetalias{Ligi2019} interferometric radius and effective temperature. The picture changes when systematic uncertainties are included, as this provides a more realistic estimate of the asteroseismic radius precision. Even then, no model achieves compatibility with the E26 radius. Interestingly, models that incorporate E26 measurements remain incompatible with E26 itself. In contrast, the situation for the \citetalias{Ligi2019} radius is markedly different: six models (those excluding E26 constraints) are compatible (we tolerated two models with $\chi^2\simeq 2.05$, which remain within rounding margins), and importantly, including those that do not rely on interferometric data.

\begin{table}[t!]
\centering
\caption{Compatibility analysis between interferometic and asteroseismic measurements of the radius of HD 219134.}
\begin{tabular}{lcccc}
\hline\hline
\multirow{2}{*}{Classical constraints} & \multicolumn{2}{c}{stat} & \multicolumn{2}{c}{stat+sys} \\ \cmidrule[0.4pt](lr){2-3} \cmidrule[0.4pt](l){4-5}
 & $\chi^2_{L19}$ & $\chi^2_{E26}$ & $\chi^2_{L19}$ & $\chi^2_{E26}$ \\
\noalign{\vskip 0.3ex}
\hline
\noalign{\vskip 0.1ex}
$R_{E26}^{deg.}$, $T_{\mathrm{eff},R21}$, $[\mathrm{Fe/H}]_{R21}$ & 3.53 & 14.78 & 2.87 & 6.51 \\
$R_{E26}^{deg.}$, $T_{\mathrm{eff},E26}$, $[\mathrm{Fe/H}]_{R21}$ & 4.29 & 12.58 & 3.51 & 5.50 \\
$R_{E26}$, $T_{\mathrm{eff},R21}$, $[\mathrm{Fe/H}]_{R21}$ & 6.55 & 7.01 & 5.48 & 3.04 \\
$R_{E26}$, $T_{\mathrm{eff},E26}$, $[\mathrm{Fe/H}]_{R21}$ & 6.77 & 6.46 & 5.66 & 2.80 \\
$R_{L19}$, $T_{\mathrm{eff},R21}$, $[\mathrm{Fe/H}]_{R21}$ & 2.09 & 23.16 & 1.69 & 10.23 \\
$R_{L19}$, $T_{\mathrm{eff},L19}$, $[\mathrm{Fe/H}]_{R21}$ & 1.64 & 27.09 & 1.33 & 11.96 \\
$L_{L25}$, $T_{\mathrm{eff},R21}$, $[\mathrm{Fe/H}]_{R21}$ & 2.54 & 18.93 & 2.05 & 8.42 \\
$L_{L25}$, $T_{\mathrm{eff},E26}$, $[\mathrm{Fe/H}]_{R21}$ & 3.60 & 14.43 & 2.93 & 6.36 \\
$L_{L25}$, $T_{\mathrm{eff},L19}$, $[\mathrm{Fe/H}]_{R21}$ & 2.55 & 18.74 & 2.05 & 8.34 \\
$T_{\mathrm{eff},R21}$, $[\mathrm{Fe/H}]_{R21}$ & 2.48 & 19.35 & 2.00 & 8.60 \\
$T_{\mathrm{eff},E26}$, $[\mathrm{Fe/H}]_{R21}$ & 3.74 & 14.04 & 3.04 & 6.18 \\
$T_{\mathrm{eff},L19}$, $[\mathrm{Fe/H}]_{R21}$ & 2.38 & 19.83 & 1.92 & 8.83 \\
\noalign{\vskip 0.3ex}
\hline
\end{tabular}
\label{tab:chi2_radius}
{\par\small\justify\textbf{Notes.} The central column (stat) is based on the statistical precision of the asteroseismic measurement of the stellar radius, while the last column (stat+sys) also includes systematic uncertainties due to grid physics and surface physics ($0.0024 R_\odot$), determined by \citetalias{Li2025} and due to the classical constraints determined in this work ($0.0019 R_\odot$). \par}
\end{table}

\begin{figure}[t!]
\centering
\includegraphics[width=0.95\linewidth]{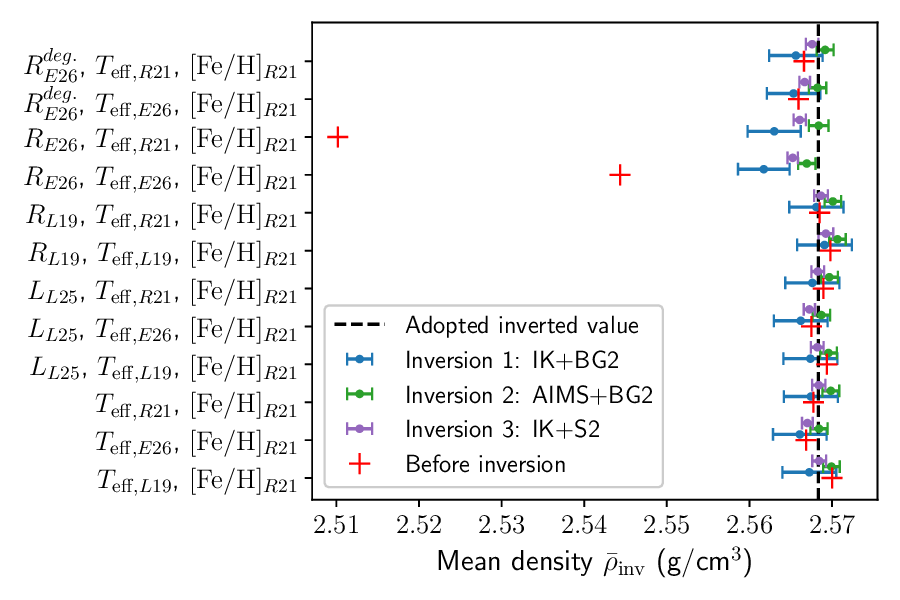}\figsep
\includegraphics[width=0.95\linewidth]{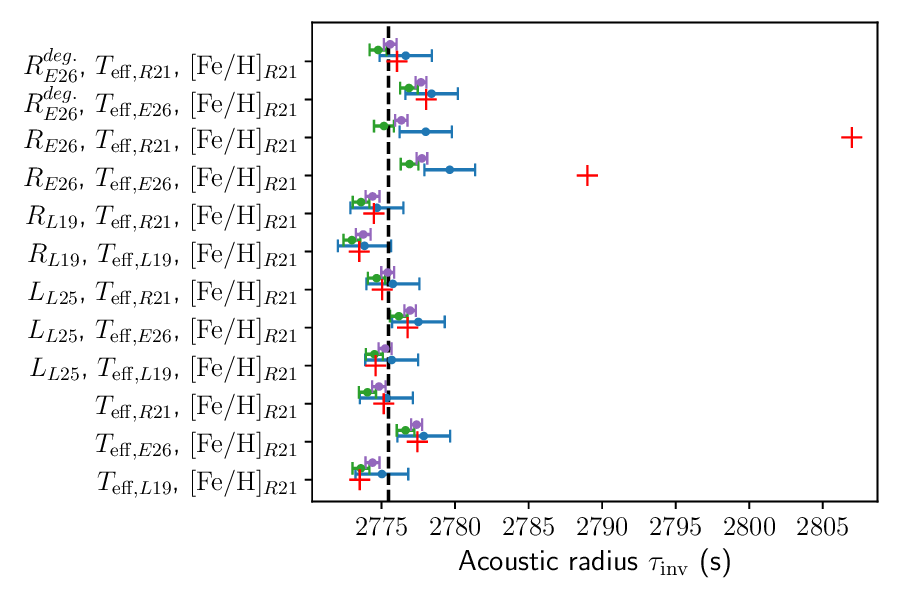}
\caption{Results from seismic inversions, showing the inverted mean density (top panel) and acoustic radius (bottom panel) for the 12 different sets of classical constraints. Three procedures were used to model the surface effects, either by optimising them during the inversion (IK+BG2 in blue and IK+S2 in violet) or using the surface effect coefficients optimised by AIMS (AIMS+BG2 in green). The red plus shows the value of the reference model on which the seismic inversion is performed.}
\label{fig:results_inversions}
\end{figure}

Figure~\ref{fig:results_inversions} displays the inversion results for the mean density and acoustic radius. Models incorporating the E26 interferometric radius as a constraint clearly deviate from the others and are rejected by the inversion, which corrects towards the solutions obtained for the remaining models. This rejection, however, involves a subtlety worth noting. Since AIMS does not interpolate the optimal stellar structure required for inversions, we recomputed the structure using the same evolutionary code version employed to generate the model grid, adopting the stellar parameters inferred by AIMS. In most cases, the recomputed structure closely matches the AIMS solution, with its frequencies remaining consistent with the interpolated AIMS frequencies \citep[see][]{Betrisey2026_catalog}. However, under certain conditions, this agreement deteriorates, and the interpolated frequencies are poorly reproduced. This is precisely what occurs for the two models in question, explaining why their recomputed mean density and acoustic radius stand out, even though the original AIMS estimates remain consistent with inversion results, unsurprising, as individual frequencies were directly fitted \citep[e.g.][]{Vandakurov1967}. In other words, AIMS interpolated a less physical solution.

Several factors can lead to such behaviour: (i) increased parameter-space degeneracy, often encountered near the end of the main sequence \citep[e.g.][]{White2011}; (ii) grid boundary effects impacting the Markov chain Monte Carlo (MCMC) sampling; or (iii) overly precise constraints that bias the sampling. HD~219134 is not near the end of the main sequence ($X_c \sim 0.44$), so the discrepancy likely stems from the latter two causes. The model grid prevents $Y_0$ from dropping significantly below physical limits, avoiding convergence on unrealistic solutions. Although the posterior distribution of $Y_0$ is moderately asymmetric, its peak lying close to the grid boundary, it is not truncated, suggesting that grid limitations had only a minor impact. As previously discussed, the angular diameter precision of E26, and thus their interferometric radius precision, is exceptionally high, improving by a factor of about three compared to \citetalias{Ligi2019}, from 2.1\% to 0.6\%. At this level, the constraint has a large weight in the cost function, strongly influencing the MCMC convergence towards a less physical solution, which is subsequently rejected by seismic inversions.

For reference, the interferometric radius is computed as $R = 0.5\theta D$, where $D$ is the distance from \citet{Bailer-Jones2021} and $\theta$ is the angular diameter. The renormalized unit weight error (RUWE) indicator is close to unity, confirming the reliability of the GAIA DR3 parallax and ruling out obvious measurement issues. Consequently, the inversion effectively rejects the angular diameter measurement underlying the E26 radius. Interestingly, the interferometric effective temperature from E26 is not rejected because it has a small weight in the cost function, but since it also depends on $\theta$, we exclude models using this quantity when estimating systematic errors due to the choice of classical constraints. These systematics are computed as the standard deviation across the six models excluding E26 measurements, yielding uncertainties of $0.006 M_\odot$, $0.0019 R_\odot$, 0.185 Gyr, $0.0004$ g/cm$^3$, and $0.005 L_\odot$, for the stellar mass, radius, age, mean density, and luminosity, respectively. For comparison, \citetalias{Li2025} report systematics of $0.014 M_\odot$, $0.004 R_\odot$, 1.0 Gyr, and $0.003$ g/cm$^3$ arising from uncertainties in physical ingredients and surface corrections.

From these results, we conclude that the E26 radius is inconsistent with asteroseismic inferences. In contrast, the \citetalias{Ligi2019} radius is compatible with asteroseismic inferences, although the seismic determination is larger by 3.4\%, a difference that is non‑negligible at the precision targeted by space‑based missions such as PLATO \citep{Rauer2025}. These findings corroborate \citetalias{Li2025}, who reported a $\sim4\sigma$ discrepancy between E26 and models excluding radius as a constraint. While \citetalias{Li2025} placed less emphasis on \citetalias{Ligi2019} measurement, their asteroseismic radius agrees with our inferences and is therefore also compatible with \citetalias{Ligi2019}. Furthermore, seismic inversions directly reject models constrained by the E26 radius \citep[see][for a recent review about seismic inversions]{Buldgen2022c}. We emphasise that this does not imply that the E26 measurement is erroneous; rather, it indicates that standard stellar models cannot reproduce it without invoking non-standard physical processes. \citetalias{Li2025} explored several such processes, including variations in mixing-length theory or atmospheric models, magnetic fields, and tidal heating, but none were sufficient to resolve the discrepancy. Given that \citetalias{Ligi2019} measurements agree with both asteroseismology and spectroscopy, however, it is reasonable to suspect that a non-trivial systematic error may have influenced the E26 angular diameter determination. While the quoted uncertainty is exceptionally small, indicating high precision, it does not necessarily ensure accuracy in the presence of systematics.

\subsection{Influence of surface modelling}
\label{sec:influcence_surface_modelling}
\citetalias{Li2025} argue that the stellar characterisation of HD~219134 has a low sensitivity to the choice of surface effect modelling, particularly for age and radius. According to their analysis, stellar ages remain consistent within uncertainties when the surface prescription is changed or when surface-dependent constraints such as $T_{\mathrm{eff}}$ are removed. Similarly, they report that stellar radius estimates are mostly unaffected by the choice of surface-effect prescription or atmospheric model, with variations of only $0.001 R_\odot$ when adopting the \citet{Sonoi2015} correction and between $0.006 R_\odot$ and $0.009 R_\odot$ depending on the tested atmospheric models.

We extended this investigation by systematically exploring all major surface effect prescriptions available in the literature \citep[see e.g.][and references therein]{Betrisey2023_AMS_surf}: \citet{Ball&Gizon2014} with one (BG1) or two (BG2) free parameters, \citet{Kjeldsen2008} with one (K1) or two (K2) free parameters, and \citet{Sonoi2015} with one (S1) or two (S2) free parameters. While BG2 and \citet{Sonoi2015} are the most widely used, it is noteworthy that Team 3 in \citetalias{Li2025} employed BG1 prescription for their characterisation. We therefore performed five additional AIMS optimisations, varying the surface effect prescription from our baseline model with BG2 (see Sect.~\ref{sec:fundamental_stellar_parameters}).

Table~\ref{tab:sensitivity_surface_physics} lists relative differences in inferred stellar mass, radius, and age introduced by each surface effect prescription with respect to our optimal model (using BG2 prescription; see Sect.~\ref{sec:fundamental_stellar_parameters}), while Fig.~\ref{fig:results_forward_and_SE} in Appendix~\ref{app:supp_data_surface_modelling} illustrates how these choices propagate into the inferred stellar parameters. Minimisations using K2 and S2 failed to converge to physical solutions, likely due to the highly non-linear nature of the second free parameter, which cannot be adequately constrained by the available seismic data. The remaining prescriptions yielded meaningful convergence, revealing two distinct behaviours. Substituting S1 for BG2 produces similar results, consistent with \citetalias{Li2025}, though we do not consider these differences negligible relative to the statistical precision (see Table~\ref{tab:sensitivity_surface_physics}) and the PLATO mission requirements for stellar age. In contrast, BG1 and K1 introduce significant changes, particularly in age. These findings agree with \citet{Nsamba2018}, who report that stellar masses of solar-like stars tend to be overestimated when using the \citet{Kjeldsen2008} and BG1 corrections, whereas BG2 and \citet{Sonoi2015} prescriptions provide the most consistent results.

Given that the \citet{Kjeldsen2008} and BG1 corrections are known to perform less efficiently due to their overly simplistic functional forms \citep[see e.g.][and references therein]{Nsamba2018,Betrisey2023_AMS_surf}, we do not consider these shifts significant for the characterisation of HD~219134. However, the fact that Team 3’s solution in \citetalias{Li2025} does not produce outliers suggests compensations with other physical ingredients, implying that the large changes observed here represent a pessimistic upper bound. Since \citetalias{Li2025} explicitly requested the five modelling teams to adopt substantially different physics, both for surface corrections and internal processes, to estimate systematic uncertainties, we regard their assessment as more realistic and robust, though it may slightly underestimate uncertainties given the limited number of physics sets considered.

Hence, the characterisation of HD~219134 is sensitive to the choice of the surface effect prescription. However, because of compensations with other physical processes, the resulting stellar parameters remain robust, with low dispersion between different modelling teams, especially for stellar mass and radius. We note that the systematics reported by \citetalias{Li2025} fold together the effects of physical-ingredient variations, surface modelling, and the usage of different modelling pipelines.

\begin{table}[t!]
\centering
\caption{Sensitivity of asteroseismic characterisation to surface physics.}
\begin{tabular}{lccc}
\hline \hline 
 & Mass & Radius & Age \\ 
\hline 
\noalign{\vskip 0.3ex}
Statistical precision BG2 & 1.9\% & 0.6\% & 10.0\% \\
\hline 
\textit{Relative differences} & & & \\
BG2 $\rightarrow$ BG1 & 3.5\% & 1.2\% & -24.7\% \\ 
BG2 $\rightarrow$ K1 & 2.4\% & 0.8\% & -17.4\% \\ 
BG2 $\rightarrow$ S1 & 0.8\% & 0.3\% & -2.7\% \\ 
\hline 
\end{tabular}
\label{tab:sensitivity_surface_physics}
{\par\small\justify\textbf{Notes.} Four surface effect prescriptions were tested: \citet{Ball&Gizon2014} with one (BG1) or two (BG2) free parameters, \citet{Kjeldsen2008} with one free parameter (K1), and \citet{Sonoi2015} with one free parameter (S1). The two latter prescriptions were also tested with two free parameters, but AIMS did not converge towards a physical solution.  \par}
\end{table}

\subsection{Fundamental stellar parameters}
\label{sec:fundamental_stellar_parameters}
Given the discrepancies in interferometric radius measurements, we opted not to include the interferometric radius as a constraint for our optimal model. Instead, our optimal model relies on acoustic oscillation data, luminosity from \citetalias{Li2025}, effective temperature from \citetalias{Ligi2019}, and metallicity from \citetalias{Rosenthal2021}. For the treatment of surface effects, we adopted the prescription of \citet{Ball&Gizon2014}, widely regarded as the most robust among existing formulations \citep{Kjeldsen2008,Ball&Gizon2014,Sonoi2015} for main-sequence and post-main-sequence solar-like oscillators \citep[e.g.][]{Ball&Gizon2017,Nsamba2018,Compton2018,Jorgensen2020,Betrisey2023_AMS_surf}. The optimal stellar parameters for HD~219134 are listed in Table~\ref{tab:stellar_characterisation}. Our model agrees closely with the median model of \citetalias{Li2025} (Team~2’s solution), but shows non-negligible differences from their optimal model (Team~1’s solution), particularly in age. This discrepancy stems from the adopted physics: \citetalias{Li2025} neglected microscopic diffusion in their optimal model, whereas we considered its inclusion essential.

\begin{table}[t!]
\centering
\caption{Optimal stellar parameters of HD 219134.}
\begin{tabular}{lc}
\hline\hline
\textit{Forward modelling} & \\
Mass $(M_\odot)$ & $0.779 \pm 0.015$ \\ 
Radius $(R_\odot)$ & $0.753 \pm 0.005$ \\ 
Age (Gyr) & $8.92 \pm 0.89$ \\ 
Mean density (g/cm$^3$) & $2.5698 \pm 0.0010$ \\ 
Luminosity $(L_\odot)$ & $0.271 \pm 0.008$ \\ 
Effective temperature (K) & $4798 \pm 41$ \\ 
$\rm [Fe/H]$ (dex) & $0.097 \pm 0.051$ \\ 
$\log g$ (dex) & $4.576 \pm 0.003$ \\ 
Initial helium abundance $Y_0$ & $0.273 \pm 0.015$ \\ 
Initial heavy-element abundance $Z_0$ & $0.018 \pm 0.002$ \\ 
\hline
\textit{Seismic inversions} & \\
Mean density (g/cm$^3$) & $2.5683 \pm 0.0028$ \\ 
Acoustic radius (s) & $2775.5 \pm 2.3$ \\
\hline
\end{tabular}
\label{tab:stellar_characterisation}
{\par\small\justify\textbf{Notes.} For the forward modelling, the quoted uncertainties correspond to the statistical uncertainty provided by AIMS based on the posterior distribution of the optimised variables. Systematics associated with the choice of the classical constraints can be found in Sect.~\ref{sec:compatibility_radii}. For seismic inversions, the quoted uncertainties also account for systematics due to the classical constraints and the choice of surface effect formalism. \par}
\end{table}

The importance of atomic diffusion in stellar interiors has been well established since the 1990s, both for the Sun \citep[e.g.][]{Proffitt&Michaud1991b,JCD1993_diffusion,Thoul1994,Basu&Anita1994,Morel1997,Vauclair1998} and for low-mass and Population~II stars \citep[e.g.][]{Proffitt&Michaud1991a,Morel&Baglin1999,Chaboyer2001,Salaris&Weiss2001}. Ignoring diffusion can introduce significant biases in asteroseismic estimates of stellar mass, radius, and age \citep[e.g.][]{Valle2014,Valle2015,Nsamba2018,Moser2023}. This explains why Team~1’s age is about 15\% larger than Team~2’s, which accounts for diffusion in their models. While K-type stars possess deeper convective envelopes that reduce the impact of diffusion on surface abundances, diffusion still alters helium and metal profiles below the convective zone in a non-negligible way. These changes affect the sound-speed profile and, consequently, asteroseismic inferences at the precision levels currently achievable. For context, PLATO’s mission requirements target age precisions of about 10\% for Sun-like stars, and we obtained a statistical precision of 10\% for the age of HD~219134. A systematic bias of 15\% from neglecting diffusion is therefore larger than the statistical precision, making its impact significant and difficult to justify neglecting microscopic diffusion for the optimal stellar model.

\citet{Takeda2007} and \citetalias{Gillon2017_HD219134} reported stellar ages of $12.46\pm 0.50$ and $11.0\pm 2.2$~Gyr, respectively, whereas our asteroseismic analysis yields a younger age of $8.92\pm 0.89$~Gyr. This agrees well with Teams~2--5 of \citetalias{Li2025}, who obtained values between 8.1 and 9.2~Gyr. Only Team~1 of \citetalias{Li2025} inferred an older age ($10.2\pm 1.5$~Gyr), again due to their neglect of microscopic diffusion, an assumption that, as discussed above, is not appropriate for HD~219134. Using stellar isochrones, \citet{Bonfanti&Gillon2020} report ages between 8.7 and 9.8~Gyr, depending on whether luminosity or interferometric radius was used as a constraint. We note, however, that this approach provides only weak constraints on age, with typical uncertainties of $\sim 30$--$50\%$ from isochrone fitting. The older age obtained by \citet{Takeda2007} is attributable to their use of an outdated version of the Yale rotational evolution code, incorporating 1990s input physics and neglecting atomic diffusion, combined with the absence of seismic information at that time. Likewise, although \citetalias{Gillon2017_HD219134} included diffusion and adopted the same abundance tables as in our work, their analysis lacked the seismic constraints required to tightly determine the stellar age, resulting in their relatively loose uncertainty. In contrast, the newly available asteroseismic data favour an age of $\sim$8.9~Gyr, or slightly younger (by about 8\%, around 8.2 Gyr) when high-metallicity abundance tables are considered.

\section{Revision of the planetary parameters}
\label{sec:revision_planetary_parameters}
Leveraging our revised stellar parameters for HD~219134, we updated the fundamental properties of planets~b, c, f, and d. Transit and orbital parameters were adopted from \citetalias{Gillon2017_HD219134}. We focused on recalculating the planetary mass $M_p$, radius $R_p$, mean density $\rho_p$, surface gravity $\log g_p$, and orbital semi-major axis $a_p$.

The planetary radii and orbital distances were obtained by rescaling the ratios $R_p/R_\star$ and $a_p/R_\star$ reported by \citetalias{Gillon2017_HD219134} using our new estimate of the stellar radius $R_\star$. The planetary masses were recomputed with the binary mass function \citep[e.g.][]{Lehmann-Filhes1894,Cumming1999},
\begin{align}
K = 	\left(\frac{2\pi G}{P}\right)^{1/3}\frac{M_p}{(M_\star+M_p)^{2/3}}\frac{\sin i}{\sqrt{1-e^2}},
\end{align}
where $K$ is the RV amplitude of an host star of mass $M_\star$, $P$ is the orbital period of the planet, $i$ is the inclination, $e$ is the orbital eccentricity, and $G$ is the Newtonian gravitational constant. Although the equation admits an analytical solution via Cardano’s method, the resulting formula is numerically unstable for the parameter ranges relevant here. We therefore determined $M_p$ using a Newton-Raphson root‑finding scheme \citep[e.g.][]{Press1986}, and propagated uncertainties through a Monte‑Carlo analysis. The planetary mean densities and surface gravities follow directly from the updated masses and radii.

All revised planetary parameters are summarised in Appendix~\ref{app:transit_orbital_planetary_parameters}. Relative to \citetalias{Gillon2017_HD219134}, our slightly lower stellar mass and radius result in correspondingly reduced planetary masses, radii, and orbital separations. The statistical precision of our stellar mass of 1.9\% represents a substantial improvement over the 3.7\% reported by \citetalias{Gillon2017_HD219134}, while the stellar radius precision remains comparable. This leads to a significant gain in the precision of the derived planetary masses. The tight radius constraint in \citetalias{Gillon2017_HD219134} originates from the availability of CHARA interferometric measurements \citep{Boyajian2012} for their study. In that case, the newly available asteroseismic data from \citetalias{Li2025} contribute modestly to the radius precision but strongly enhance the constraints on the stellar mass and age. For HD~219134, systematic uncertainties in the stellar parameters are very significant compared to the statistical ones (Sect.~\ref{sec:compatibility_radii}). Our final uncertainties therefore appear slightly larger, as we account for sources of error previously neglected. Finally, our inferred planetary properties agree well with those reported by \citetalias{Li2025} for HD~219134b and c.

\section{Host-star rotational history and planetary mass loss}
\label{sec:stellar_rotation_and_planet_mass_loss}

The distribution of surface rotation rates of low-mass, solar-like stars spans a wide range of values during the early evolutionary stages, globally reaching their maxima at the zero-age main sequence, and finally converging into a `slow rotators' sequence approximately half-way through the main sequence. Retrieving the rotational histories of these stars when they are on the main sequence is challenging, since a strong degeneracy dominates at earlier stages. The rotational histories of stars have a significant impact on the efficiency of tidal and radiative interactions with close-by orbiting planets. In this section, we study the evolution of the surface rotation rate of HD~219134, and the potential impact of tidal and radiative interactions on confirmed orbiting planets.

\subsection{Instantaneous angular momentum loss from 3D stellar wind models}
\label{Sect:AM_loss_ZDI_3D}
To calculate the present-day angular momentum loss rate of HD 219134, we employed a 3D magnetohydrodynamics model for the stellar wind, in which the assumed surface magnetic field of the star is derived from spectropolarimetric observations and reconstructed using Zeeman-Doppler imaging (ZDI). For the reconstructed surface magnetic field, we used the ZDI map presented in \citet{Folsom2018}. The mass, radius and rotation rate of the star assumed in our model is the same as derived in our current study.

The stellar wind model is similar to that presented in other works in the literature \citep{Vidotto2023,Evensberget2023,Bellotti2026} and is based on the model developed for studying the solar wind \citep{vanderHolst2014}. In our wind model, the dissipation and propagation of Alfven waves provide the energy and momentum to drive the stellar outflow, contrary to the model presented in our previous study of HD 219134 \citep{Vidotto2018}, where the acceleration of the wind was essentially provided by thermal forces. Both of these models have free parameters: while in thermally driven wind models the base wind density and temperature are unknown (and the solution is rather sensitive to these unknowns), in Alfven-wave driven wind models, the flux of Alfven waves is the main unknown \citep{vidotto2021}. To 
constrain the Alfven wave flux, we compared the resulting mass-loss rates from our models with the values derived in \citet{Folsom2018}, which were based on astrospheric Ly-$\alpha$ observations \citep[as in e.g.][]{Wood2021}. In that work,  the mass-loss rate of HD 219134 was constrained to be in the range between $1\times 10^{-14}$ and $2 \times 10^{-14} M_\odot$~yr$^{-1}$, i.e. between half to one times the solar wind mass-loss rate. 

In practice, to constrain the wave flux, we ran three stellar wind models with fluxes $S_A/B = [0.1, 1, 10] \times 1.1 \times 10^{5}$~erg~(s cm$^2$ G)$^{-1}$. We note that the fluxes quoted before were divided by the local value of the stellar magnetic field $B$. This implies in practice that, while the energy density of the waves in the stellar surface is homogeneous, the wave fluxes are higher (lower) in regions of higher (lower) field strength. This is described in more details in \citet{Vidotto2023}. For comparison, the solar wind models assume a value of $S_A/B = 1.1 \times 10^{5}$~erg~(s cm$^2$ G)$^{-1}$, implying that the models we ran for HD 219134 span fluxes ten times smaller and larger than those typically assumed in solar wind models. Our three models yield the following mass-loss rates $[6.8\times 10^{-16}, 9.2\times 10^{-15}, 2.0 \times 10^{-13}]\, M_\odot$~yr$^{-1}$ and the following instantaneous angular momentum-loss rates $[3.1 \times 10^{28}, 1.1 \times 10^{29}, 6.5 \times 10^{29}]$~erg. Comparing to the observed mass-loss rates value, we constrain the wave flux of HD 219134 to be  $S_A/B \simeq  [1.1$ to $2.2] \times 10^{5}$~erg~(s cm$^2$ G)$^{-1}$. With this, our derived angular momentum-loss rates are  $\simeq [1.1$ to $1.9]\times 10^{29}$~erg. Another quantity that can be derived from the solar wind models is the position of the Alfven surface that we provide in Appendix~\ref{app:position_Alfen_surface} for completeness.

\subsection{Rotational history}
\label{sec:rotational_history}

To reconstruct the rotational evolution of HD~219134, we provided the stellar structural track computed with CLES, on the basis of the optimal fundamental parameters determined above, to a star-planet interaction code \citep{Privitera2016_spi2, Rao2018, Pezzotti2021_K444, Pezzotti2025}. By means of this code, the evolution of the host star angular momentum is computed under the assumption of solid body rotation, accounting for the contribution of two main mechanisms (beside the variation of the momentum of inertia): the braking at the surface by magnetised stellar winds, and the exchange of angular momentum between the planetary orbit and the stellar surface due to tidal interaction with close-in, massive planetary companions. The first mechanism is included following the prescription of \citet{Matt2015, Matt2019}, which accounts for the loss of angular momentum in the so-called unsaturated and saturated regimes \citep[e.g.][]{Pizzolato2003, Wright2011}. The transition between these two regimes is regulated by the stellar Rossby number, defined as the ratio between the surface rotation period and the convective turnover timescale ($Ro = P_{\rm rot}/\tau_{\rm conv}$): stars with $Ro \leq Ro_{\rm sat}$ are in the unsaturated regime, while stars with $Ro > Ro_{\rm sat}$ are in the saturated one, with $Ro_{\rm sat} = Ro_{\odot}/\chi$, where $Ro_{\odot}$ is the solar Rossby number and $\chi$ is a parameter calibrated by means of activity-rotation relationships \citep[e.g.][]{Wright2011, Johnstone2021}. In this work we considered $\rm \chi = 13$, as obtained from calibration of the X-ray luminosity versus Rossby number relationship for low-mass stars as in \citet{Pezzotti2021_K444}.

For stars more evolved than the Sun, with $Ro \gtrsim Ro_{\rm cri}$ and $Ro_{\rm cri} = 0.92~Ro_{\odot}$ \citep{Metcalfe2024}, it is possible to account for the onset of a weakened-magnetic-braking (WMB) regime \citep{VanSaders2016}: once stars enter this regime, the efficiency of the braking at their surface is hampered by a shift in the morphology of the magnetic field, from larger to smaller spatial scales \citep{Reville2015,Garraffo2016}, and (or) by an abrupt change in the mass-loss rate \citep{OFionnagain2018}. In this work, we account for this effect by considering an exponential decay of the angular momentum loss with time. According to our models, HD~219134 would enter the WMB regime at $t_{\rm wmb} \sim 7.49$ Gyr.

The system of equations for the angular momentum loss in the three regimes mentioned above reads:
\begin{equation}\label{eq:ΜΒ_WMB}
\dot{J}_{\rm wind} =
\begin{cases}
T_0 ~\chi^{\rm p} \left( \dfrac{\Omega_{\star}}{\Omega_{\odot}} \right), &\textit{ if} ~~ Ro \leq\dfrac{Ro_{\odot}}{\chi},\\
 T_0 \left(\dfrac{\tau_{\rm conv}}{\tau_{\rm conv \odot}} \right)^{\rm p} \left(  \rm \dfrac{\Omega_{\star}}{\Omega_{\odot}} \right)^{\rm p+1}, &\textit{ if}~~ \dfrac{Ro_{\odot}}{\chi} < Ro <  Ro_{\rm cri},\\
\dot{J}_{\rm (wind, 0)}~\mathrm{e^{-\lambda (t - t_{\rm wmb})}}\quad, &\textit{ if} ~~ Ro_{\rm cri} \leq Ro, 
\end{cases}
\end{equation}
where $\rm \tau_{conv}$ is the convective turnover timescale of the star at the time $t$ \citep{Rasio1996, Villaver2009}, $\tau_{\rm conv \odot}$ is the convective turnover timescale of the Sun, and $\rm Ro_{\odot}$ is the solar Rossby number. The exponent $\rm p = 2.1$ and the braking constant $T_{\odot} = 8 \times 10^{30}~\rm erg$ in the expression $T_0 = -T_{\odot} \left({R_{\star}}/{R_{\odot}} \right)^{3.1} \left({M_{\star}}/{M_{\odot}} \right)^{0.5}$ are typically calibrated to reproduce the surface-rotation rate of the Sun at solar age ($\Omega_{\odot} = 2.9 \times 10^{-6}~\rm s^{-1}$). The quantity $\dot{J}_{\rm (wind, 0)}$ is the stellar angular momentum loss at $Ro = Ro_{\rm cri}$, and the parameter $\lambda = 1.98 \times 10^{-9}~\rm Gyr^{-1}$ has been calibrated to reproduce the $\dot{J}_{\rm wind, ZDI} = 1.1 \times 10^{29}$~erg derived in Sect.~\ref{Sect:AM_loss_ZDI_3D} at the stellar age, under the assumption\footnote{Adopting $\dot{M} = 1~\dot{M}_{\odot}$ would yield similar results.} that the star is experiencing a mass loss rate $\dot{M}=0.5\dot{M}_{\odot}$ \citep{Folsom2018}.

\begin{figure}[t]
\centering
\begin{subfigure}{
\includegraphics[width=0.93\linewidth]{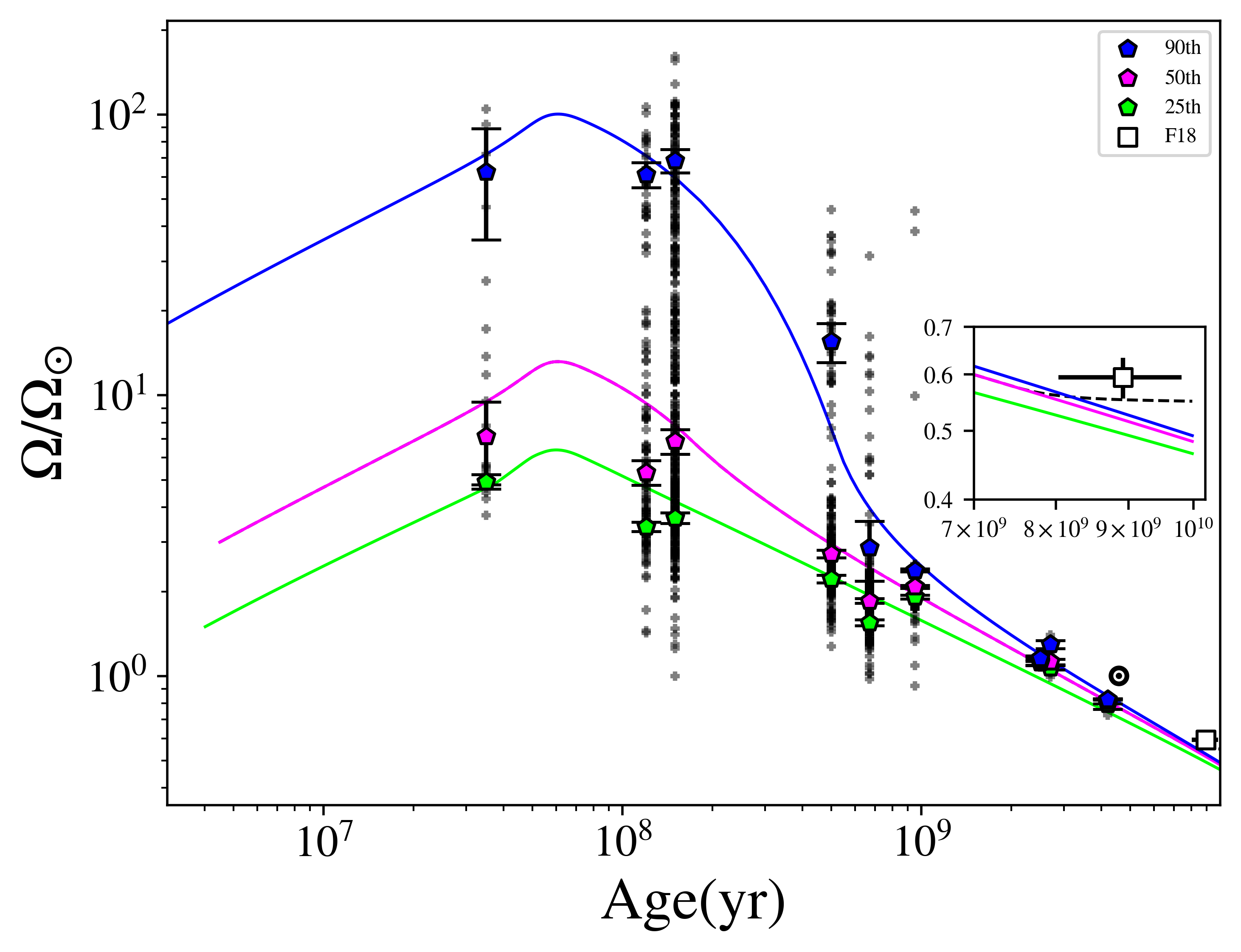}
}
\end{subfigure}
\vspace{-1.0\baselineskip} 
\par 
\begin{subfigure}{
\includegraphics[width=0.955\linewidth]{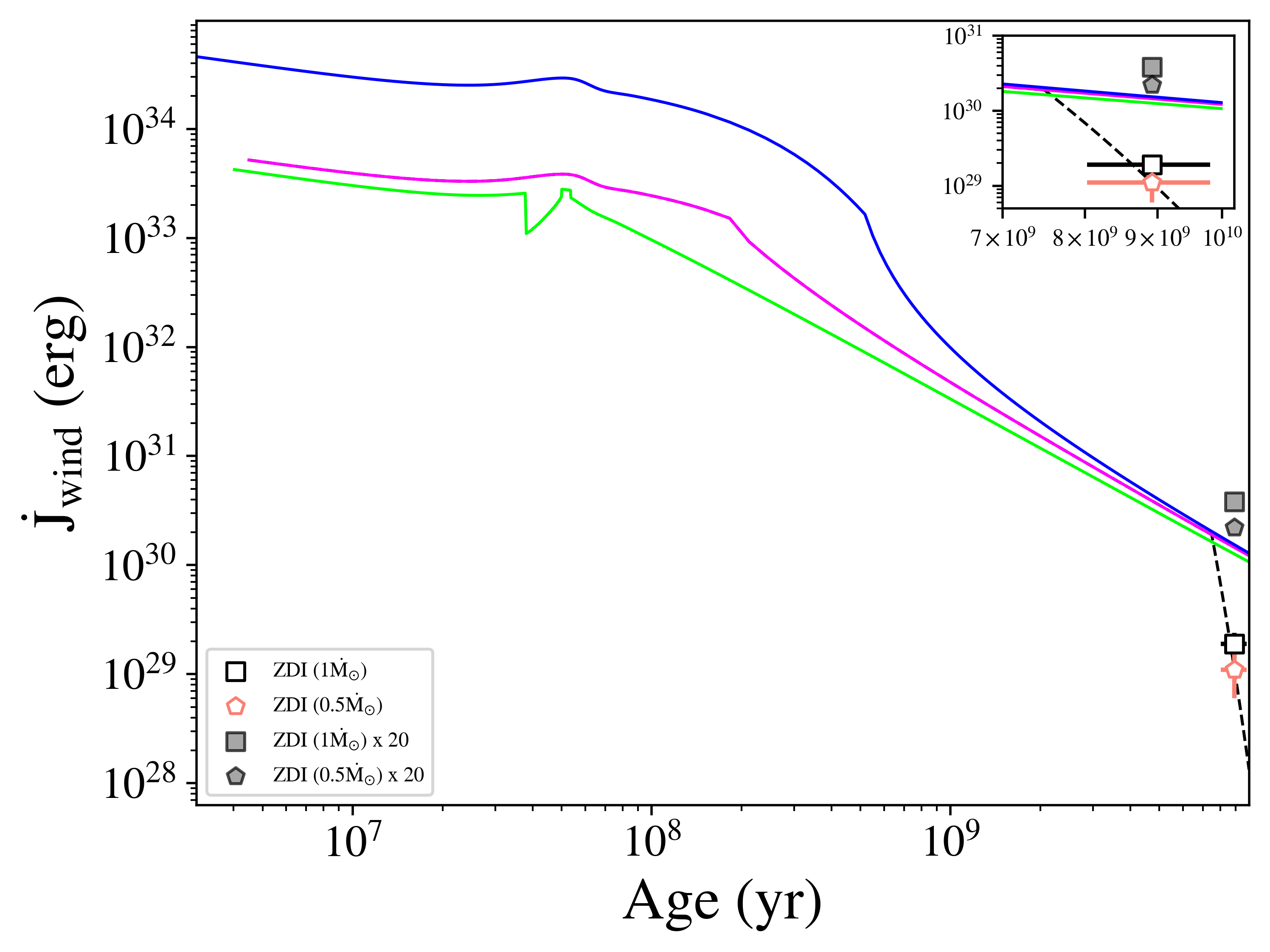}
}
\end{subfigure}
\captionsetup{skip=0.0pt}
\caption{\textit{Top panel:} Evolution of the surface rotation rate, normalised to the solar value, for three initial surface rotation rates, $\Omega_{\mathrm{in}}/\Omega_{\odot} = 3.2, 5.0, 18$, corresponding to slow (green), moderate (magenta), and fast rotator (blue). The square shows the observed rotation rate from \citet{Folsom2018}. The black crosses show surface rotation rates for stars in open clusters, as detailed in Appendix~\ref{app:calibration_braking_law}. \textit{Bottom panel:} Evolution of the angular momentum loss due to magnetised winds for the same rotational histories. The salmon pentagon and the black square show  $\dot{J}_{\mathrm{wind}}$ from 3D stellar wind models assuming mass-loss rates of $\dot{M}(\dot{M}_{\odot}) = 0.5$ and $1$, respectively. The same symbols in grey indicate the rescaled values, $\dot{J}_{\mathrm{wind}}\times 20$, at the system age.}
\label{Fig:rot_Jdot}
\end{figure}

Given the significant difference in spectral type between HD~219134 and the Sun, we used a dedicated calibration of the braking law in the unsaturated-saturated regimes based on observational data from open clusters with ages between 35~Myr and 4.2~Gyr (see Appendix~\ref{app:calibration_braking_law}). In the top panel of Fig.~\ref{Fig:rot_Jdot}, the 90th, 50th and 25th percentiles for each cluster are represented by the blue, magenta and green pentagons, respectively. The corresponding tracks for the fast, moderate and slow rotator are indicated by the solid lines, with the same colour code. The dashed black line shows the evolution of the surface rotation rate subsequently to the onset of WMB. By comparing the theoretical tracks without WMB with HD~219134's measured rotation rate ($\Omega_{\rm HD219134} = (0.594 \pm 0.039)~\rm s^{-1}$, \citealt{Folsom2018}), indicated by the square symbol, we notice that they tend to slightly underestimate it ($\Omega_{\rm fast} = 0.53\rm~ s^{-1}$), remaining compatible within two sigma. The dashed black track instead appears to be compatible within one sigma. In the bottom panel of Fig.~\ref{Fig:rot_Jdot}, we show the angular momentum loss tracks for each of the considered rotators. The black square and salmon pentagon indicate the $\dot{J}_{\rm wind}$ estimates determined in Sect.~\ref{Sect:AM_loss_ZDI_3D}. In \citet{Bellotti2026}, using rotational tracks computed with the braking law from \citet{Matt2015, Matt2019} for the Sun with the same calibration as in \citet{Pezzotti2025}, the authors found that a factor of 20 (similarly to \citealt{Finley2019b}) is needed to reconcile the rotational tracks at the solar age with the torque determined at the solar maximum. If we multiply the $\dot{J}_{\rm wind}$ derived in Sect.~\ref{Sect:AM_loss_ZDI_3D} by a factor of 20 (showed by the grey square and pentagon), we notice that a better agreement is retrieved with the theoretical tracks (without WMB). In general, we note that a comparison between the angular momentum loss inferred from rotational evolution tracks (without WMB) and the semi-empirical values derived from ZDI observations requires the application of an appropriate scaling factor \citep{Finley2019b}. The dashed black line shows the trend of the angular momentum loss after the WMB sets in.

By comparing the theoretical tracks with the observational constraints on the surface rotation rate and angular momentum loss, we find that tracks without WMB are consistent within two sigma with the observed surface rotation rate and are compatible with the rescaled $\dot{J}_{\rm wind}$, in agreement with results obtained for other solar-like stars \citep{Finley2019b, Evensberget2023}. Conversely, tracks including WMB are consistent within one sigma with the observed surface rotation rate when adopting a WMB prescription calibrated on the semi-empirically derived $\dot{J}_{\rm wind}$. In this case, however, the rescaling between the angular momentum loss inferred from rotational tracks without WMB and that obtained semi-empirically from ZDI maps would no longer apply. In this respect, new observations particularly targeting stars more evolved than the Sun are essential to determine whether the rescaling between the theoretically inferred and the semi-empirically derived $\dot{J}_{\rm wind}$ still applies for more evolved objects.

The tidal interaction with close-in planets is accounted for by means of equilibrium and dynamical tides \citep[through the excitation of inertial waves;][]{Ogilvie2013} dissipated within the host star convective envelope \citep[][and references therein]{Privitera2016_spi1, Privitera2016_spi2, Rao2018} by assuming that the planet is on a circular-coplanar orbit around the host star. This tidal interaction potentially leads to either outward or inward migration of planets and to either spin-down or spin-up of the host star. In the specific case of HD~219134, assuming that the planets reached their current orbital distance at the dispersal of the protoplanetary disk, we do not expect tides to have played a crucial role in determining the consequent evolution of their orbits. Despite the system’s compact architecture, the masses of HD~219134b and c are insufficient to drive significant tidal interactions with the host star. Similarly, the two outermost planets (HD~219134d, and h) although comparatively more massive than the inner ones 
(see Table~\ref{tab:planetary_parameters})
are located at too large orbital distances to produce substantial tidal dissipation in the star, at least within the framework of the tidal mechanisms implemented in our model \citep{Pezzotti2021_K444, Pezzotti2025}. The evolution of the host star surface rotation rate on the main sequence is thus driven by the mass loss via magnetised stellar winds.

\subsection{Stellar X-ray luminosity and planetary atmospheric mass loss}
\label{sec:xray_luminosity_and_atmospheric_mass_loss}

\begin{figure}[t]
    \centering
\includegraphics[width=\linewidth]{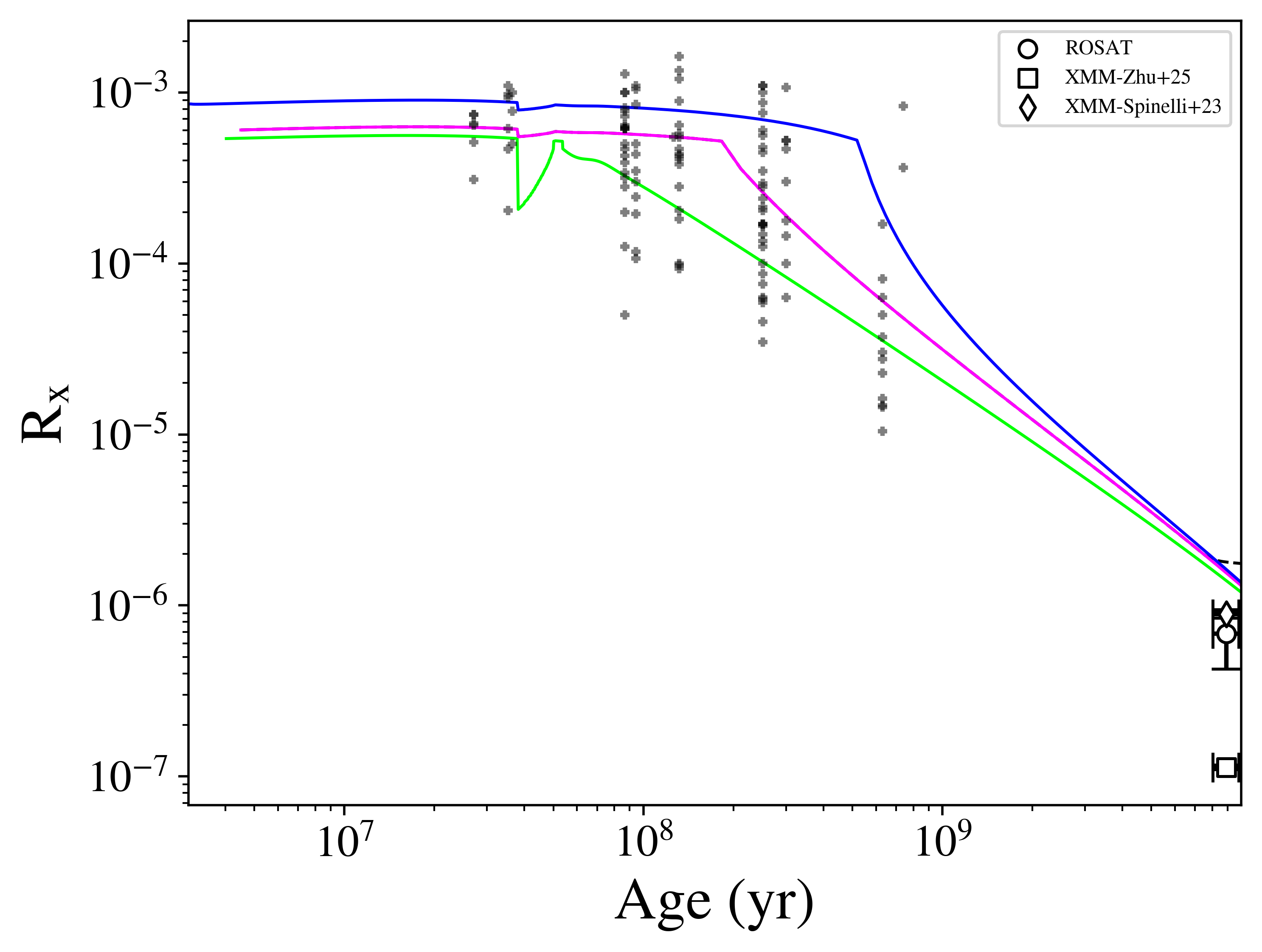} 
    \caption{Evolution of the X‑ray‑to‑bolometric luminosity ratio for fast (blue), intermediate (magenta), and slow (green) rotators. The black symbols show the X-ray luminosity from observations with ROSAT (circle) and XMM-Newton (diamond and square, with the latter converted to the ROSAT energy band). The grey crosses show ROSAT measurements of K‑type stars in open clusters for comparison \citep{Jackson2012, Pezzotti2026}.}
    \label{Fig:Lx}
\end{figure}

The star-planet interaction code models atmospheric mass loss from H-rich planetary atmospheres induced by stellar X-ray and extreme ultraviolet (EUV) irradiation. The evolution of the stellar X-ray luminosity is computed self-consistently with the one of the surface rotation rate via the $R\rm x$–$Ro$ relationship \citep[e.g.][]{Wright2011, Johnstone2021, Pezzotti2021_K444, Pezzotti2026}, where $R\rm x$ is the X-ray-to-bolometric luminosity ratio and $Ro$ the stellar Rossby number. In this work, we adopted the $R\rm x$–$Ro$ recalibrated from \citet{Pezzotti2021_K444}. The EUV luminosity is subsequently computed by rescaling the X-ray one as in \citet{Johnstone2021}. The theoretical tracks for the fast, moderate, and slow rotators derived using this relation are shown by the solid blue, magenta, and green lines in Fig.~\ref{Fig:Lx}. The dashed black line represents the evolution of $R \rm x$ in the WMB regime, assuming that the $R\rm x$–$Ro$ relation described above remains valid. The diamond shows the estimate derived in \cite{Spinelli2023} using XMM-Newton.  The circle and square symbols indicate the observationally derived X-ray luminosities of the star reported by \citet{Zhu2025}: the circle corresponds to observations performed with ROSAT, while the square corresponds to observations obtained from XMM-Newton after conversion into the ROSAT energy band (0.1–2.4 keV). We notice that the observations performed with ROSAT (in 1991) and XMM-Newton (in 2016) are related to slightly different phase moments of the magnetic cycle on the rising branch towards the maximum (see Fig.~\ref{Fig:S-index_Lx} in Appendix~\ref{app:magnetic_cycle}). It is worth noting that the X-ray luminosity estimates obtained by \citet{Zhu2025} and \citet{Spinelli2023} from XMM-Newton data differ by approximately a factor of 7. This might be due to a different approach in the analysis of the X-ray spectrum performed in these studies. 

In \citet{Pezzotti2026}, we investigated the relationship between X-ray luminosity and stellar age for a sample of asteroseismically well-characterised solar-like stars, with the goal of probing the evolution of mean stellar activity and the underlying dynamo mechanism at advanced evolutionary stages. In Appendix~\ref{app:X_age_relationship}, we present the location of HD~219134 in the $\rm Log(L_x/R^2)-Log(Age)$ plane relative to the stars analysed in \citet{Pezzotti2026}. HD~219134 is the oldest star in the sample and appears to follow the observed trend of decreasing activity with increasing age. Long-term monitoring of HD~219134 in the X-ray would be highly desirable to test whether a significant change in the X-ray flux is observed at different phases of the magnetic activity cycle and better understand the dynamos in very old stars.

\begin{figure}[t]
    \centering
\includegraphics[width=\linewidth]{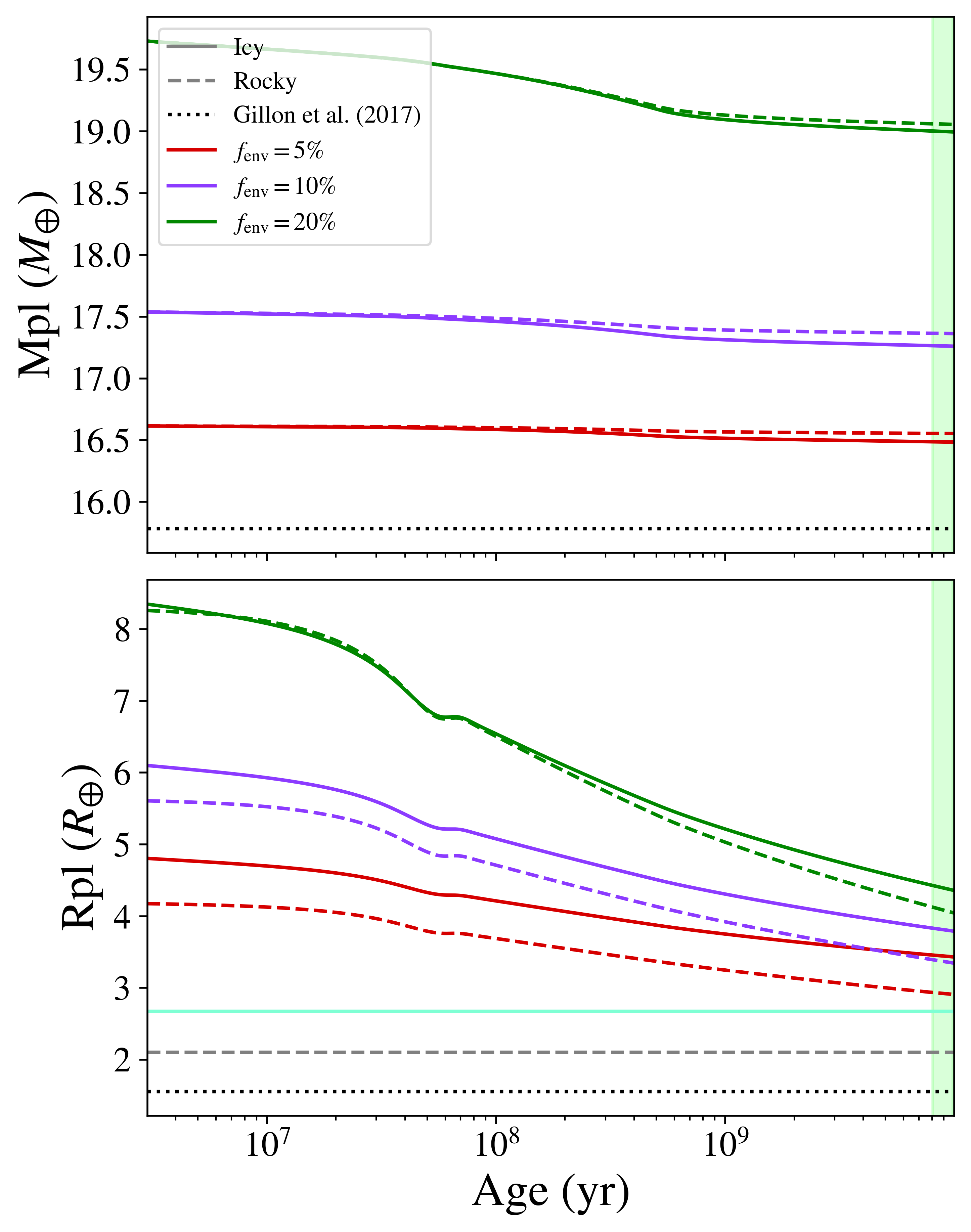} 
    \caption{\textit{Top panel:} Evolution of the total mass of HD~219134d assuming an initial icy (solid lines) or rocky (dashed lines) core mass of $15.784~M_{\oplus}$ and H-rich envelope fractions $f_{\mathrm{env}} = 5\%$ (red), 10\% (violet), and 20\% (green). The black dotted line indicates the assumed core mass (cf. Table~\ref{tab:planetary_parameters}). The light-green vertical band shows the age uncertainty. \textit{Bottom panel:} Evolution of the planetary radius for the two core compositions and the envelope mass fractions as indicated above. The light‑blue solid line and grey dashed line denote the icy and rocky core radius, respectively. The black dotted line shows the estimate for the core radius as in Table~\ref{tab:planetary_parameters}.}
    \label{Fig:HD219134_evaporation}
\end{figure}

The mass loss from the planetary atmosphere is estimated by interpolating or extrapolating on the grid of upper-atmosphere models of \citet{Kubyshkina2018, Kubyshkina2021} at each time-step of the simulation, providing the EUV flux as input. For what concerns the atmospheric mass loss for the two innermost planets, given their proximity to the host star and relatively low gravity, any primordial H-He rich atmosphere has been stripped within a few millions years from the dispersal of the protoplanetary disk. The two outermost, non-transiting, confirmed planets instead (HD~219134d, HD~219134h~\footnote{\url{https://exoplanetarchive.ipac.caltech.edu/}}), could have retained a fraction of the primordial atmosphere thanks to their larger gravity and lower XUV-flux irradiation. In particular, the most external planet, HD~219134h, with a lower limit on mass $M_{\rm p}\mathrm{sin(i)}(M_{\oplus}) = 97.9 \pm 4.4$ and orbit $a\mathrm{(AU)} = 2.968 \pm 0.037$ \citep{Rosenthal2021}, is the most resilient to evaporation of the atmosphere. Planet `d' instead could have significantly undergone higher mass loss. To obtain an estimate of the lifetime mass loss experienced by HD~219134d, we simulated the evolution of the system by assuming that the star evolved as a fast rotator (to maximise the impact of XUV-driven mass loss), and that the planet occupied its current orbit already at the dispersal of the protoplanetary disk. We also assumed that it formed with core mass $M_{\rm core}(M_{\oplus}) = 15.784$ (see Table~\ref{tab:planetary_parameters}) and varied the initial H-rich atmospheric mass fraction, considering $f_{\rm env} = 5\%, 10\%, 20\%$. For the computation of the planetary radius, we adopted the fitting formulae of \citet{Chen2016} and tested the mass-loss from the planet under the assumption of icy or rocky composition of the core. The results are reported in Fig.~\ref{Fig:HD219134_evaporation}, where in the top panel we show the evolution of the total planetary mass for the diverse initial atmospheric mass fractions and in the bottom panel the corresponding evolution of the total radius for the two assumed compositions. Concerning the evolution of the total mass, we note that the XUV-driven mass loss has reduced the planetary mass by only a few percent of its initial value at the age of the system, with significantly stronger impact in the case $f_{\rm ini} = 20\%$. The evolution of the total planetary radius is dominated by the cooling of the planet, rather than the XUV-driven mass loss. All tracks converge in a range of planetary radii between 3 and 4.5~$R_{\oplus}$. We notice that the core radius derived by the fitting formulae of \citet{Chen2016} in this work is larger than the one estimated in \citet{Gillon2017_HD219134} and the one therefore revised in Sec.~\ref{sec:revision_planetary_parameters}, for both the considered compositions.

The lack of more stringent constraints on both the planetary orbital inclination and the planetary radius (or density) makes the problem highly degenerate. Therefore, although these results should be regarded as indicative in this context, they can still provide a reasonable estimate of the impact of XUV-driven mass loss, and thus of the star’s average activity on the planet, with potential implications for its composition.

\section{Conclusions}
\label{sec:conclusions}
We revisited the stellar characterisation of HD~219134, employing complementary seismic diagnostics and advanced modelling techniques, as detailed in Sect.~\ref{sec:asteroseimic_modelling}. These efforts allowed us to refine the stellar parameters and, in turn, to update the planetary properties in Sect.~\ref{sec:revision_planetary_parameters} and to evaluate the atmospheric mass loss experienced by the system's close‑in planets in Sect.~\ref{sec:stellar_rotation_and_planet_mass_loss}.

\citetalias{Li2025} highlight a significant tension between the asteroseismic radius inferred for HD~219134 and the recent interferometric determination by E26. Our complementary analysis, combining both forward modelling and seismic inversions, confirms that the E26 radius is incompatible with the standard physics implemented in our asteroseismic models. Notably, our conclusions are firmer than those reached by \citetalias{Li2025}. While their argument was indirect, showing that omitting the E26 radius leads to a different seismic solution, we demonstrated that the E26 measurement is directly rejected by seismic inversions.

We also assessed the compatibility between our seismic results and the earlier interferometric measurement of \citetalias{Ligi2019}. Once systematic stellar uncertainties are included, the \citetalias{Ligi2019} radius remains consistent with our seismic predictions, although our preferred seismic radius is 3.4\% larger, which is not negligible at the precision level of targeted by missions such as PLATO, which requires 2\% accuracy for the radius of a Sun-like star. Furthermore, the effective temperature derived by \citetalias{Ligi2019} is in better agreement with the spectroscopic analysis of \citetalias{Rosenthal2021}. Recent independent interferometric results (Nowacki et al., in prep.) yield a radius fully consistent with our asteroseismic determination and suggest that the uncertainties reported in all previous interferometric studies may have been significantly underestimated. Together with our seismic analysis, these results suggest that the apparent radius discrepancy for HD~219134 does not reflect a genuine astrophysical problem. Instead, it is more likely explained by underestimated uncertainties in previous interferometric analyses, particularly E26. Future independent measurements and cross-validation between interferometric teams will nevertheless be valuable to fully resolve the origin of these systematics and consolidate HD~219134 as a benchmark target for high-precision stellar physics of K~dwarfs.

In addition, we show that the treatment of the surface term can have a significant impact on the seismic modelling of HD~219134, contrary to the conclusions of \citetalias{Li2025}. Interestingly, their finding of little impact of surface modelling results from compensations between the surface treatment and simultaneous variations of the micro- and macro-physical ingredients. This interplay remained hidden in their analysis because both sources of bias were varied concurrently, allowing degeneracies to mask their respective contributions. As a consequence, the systematic uncertainties reported by \citetalias{Li2025} fold together the effects of micro- and macro-physical ingredient variations and surface modelling, as well as the usage of different modelling codes.

We obtain revised stellar mass, radius, and age of $0.779 \pm 0.015_{\mathrm{(stat)}} \pm 0.006_{\mathrm{(sys)}}\, M_\odot$, $0.753 \pm 0.005_{\mathrm{(stat)}} \pm 0.002_{\mathrm{(sys)}}\, R_\odot$, and $8.92 \pm 0.89_{\mathrm{(stat)}} \pm 0.19_{\mathrm{(sys)}}~\mathrm{Gyr}$, respectively. Stellar mass and radius are within the quality requirements targeted by PLATO \citep[15\% in mass, 2\% in radius, and 10\% in age;][]{Rauer2025}, but stellar age exceeds them if systematics are included. Our inferred age is younger by $\sim$1.2 Gyr compared with that from \citetalias{Li2025}, a difference that stems from our treatment of atomic diffusion. We did not adopt the assumptions of \citetalias{Li2025} as we do not believe they are applicable for HD~219134, as outlined in Sect.~\ref{sec:fundamental_stellar_parameters}. Notably, our age estimate is in good agreement with the results of Teams~2 (their median model) and~3 in \citetalias{Li2025}, which apply a diffusion treatment similar to ours. Using an appropriate diffusion formalism yields seismic ages around $\sim$8.9 Gyr, or slightly younger for high-metallicity abundances (Teams~4 and~5 in \citetalias{Li2025}). In complement to \citetalias{Li2025}, we quantified the systematic uncertainties associated with the non-seismic constraints. These amount to $0.006\,M_\odot$, $0.0019\,R_\odot$, $0.185$ Gyr, $0.0004\,\mathrm{g\,cm^{-3}}$, and $0.005\,L_\odot$ for the stellar mass, radius, age, mean density, and luminosity, respectively. For reference, \citetalias{Li2025} report systematic uncertainties of $0.007\,M_\odot$, $0.002\,R_\odot$, $0.81$~Gyr, and $0.003\,\mathrm{g\,cm^{-3}}$, arising from combined uncertainties arising from physical ingredients and surface-effect corrections. The resulting revised stellar properties establish HD~219134 as a robust benchmark for testing rotational evolution and magnetic braking in K dwarfs, and we build on this foundation to investigate its rotational and magnetic evolution and interaction with the planetary system using stellar evolutionary models, rotational evolution calculations, stellar wind constraints, and atmospheric escape simulations.

We computed 3D stellar wind models to derive the present‑day angular momentum‑loss rate. Our models used the reconstructed stellar magnetic field from \citet{Folsom2018}, with free parameters calibrated to reproduce their inferred mass‑loss rate of $[1$ to $2]\times 10^{-14}M_\odot$~yr$^{-1}$. This yields a present‑day angular momentum‑loss rate of $\simeq[1.1$–$1.9]\times10^{29}$~erg, which we used to anchor the rotational evolution of HD~219134.

Using a dedicated calibration of the magnetic braking law for low‑mass stars based on open‑cluster observations, we reconstructed the rotational history of the host star and compared the predicted evolutionary tracks with constraints on the present‑day surface rotation and angular momentum loss. Standard magnetic braking prescriptions reproduce the observed rotation rate within two sigma and remain compatible with the rescaled angular momentum‑loss rate inferred from the stellar wind models \citep{Finley2019b, Evensberget2023, Bellotti2026}. Models invoking weakened magnetic braking yield an even closer match to the observed rotation at the one‑sigma level when calibrated on the semi‑empirical estimate of $\dot{J}_{\rm wind}$. However, in this regime the commonly adopted scaling relations used to reconcile theoretical and observationally inferred angular momentum‑loss rates cease to apply. This tension emphasises the need for improved observational constraints on stellar winds and magnetic braking in stars more evolved than the Sun to clarify both the onset and efficiency of the WMB regime \citep[e.g.][]{VanSaders2016, Metcalfe2024, Metcalfe2026}.

We further examined the evolution of the stellar X‑ray luminosity and its consequences for planetary atmospheric erosion. The reconstructed activity history implies that the two innermost planets were likely stripped of any primordial H–He envelopes at early times owing to intense XUV irradiation. In contrast, the more massive and wider‑orbit planets HD~219134d and~h are expected to have retained at least a fraction of their primordial atmospheres \citep{Kubyshkina2018}. In particular, HD~219134d appears to have undergone only moderate atmospheric erosion over the system lifetime, even assuming an initially rapid stellar rotation and high activity level. For this planet, the radius evolution is dominated by thermal contraction rather than by atmospheric escape. Taken together, these results establish HD~219134 as a benchmark system for probing the long‑term rotational evolution of K dwarfs, the potential transition towards WMB, and the cumulative impact of stellar activity on planetary atmospheres. Future long‑baseline X‑ray monitoring, improved constraints on stellar winds, and more precise determinations of planetary properties, especially for the outer planets, will be crucial to further refine the evolutionary picture of this system.

\begin{acknowledgements}
The authors thank Dr. Daniel Reese for his valuable assistance in defining the compatibility indicator used in this work. J.B. acknowledges funding from the SNF Postdoc.Mobility grants no. P500PT{\_}222217 and P5R5-2{\_}239064. C.P. has received funding from the Italian Space Agency (ASI) and the Ministry of University and Research (MUR) through ‘Space It up’ Spoke~6. G.B. acknowledges funding from the Fonds National de la Recherche Scientifique (FNRS) as a postdoctoral researcher. A.A.V. acknowledges support from the Dutch Research Council (NWO) through project No. VI.C.232.041 (Talent Programme Vici). C.P.F. acknowledges funding from the Estonian Research Council grant TARISTU24-TK3 under ESA Estonia research infrastructure. M.G. is a FNRS Research Director.
\end{acknowledgements}


\bibliography{bibliography.bib}

\appendix

\section{Compatibility indicator between two measurements}
\label{app:compatibility_indicator}
Let us assume we have a given quantity (in our case, the true radius of the star), and obtain two probability distributions
for this quantity (in our case, from interferometry and from asteroseismology). We want to define an indicator that assesses the compatibility of these two distributions. For the sake of simplicity, we assume these distributions
to be independent and Gaussian.

We have the following estimates: $v_A \pm \sigma_A$ and $v_B \pm \sigma_B$. We want to choose a value of $v$ that is the closest possible to these two estimates, which can be achieved by reducing the following $\chi^2$~value:
\begin{equation}
\chi^2 = \left( \frac{v - v_A}{\sigma_A} \right)^2
       + \left( \frac{v - v_B}{\sigma_B} \right)^2.
\end{equation} 

Setting $\frac{d\chi^2}{dv}=0$, we get:
\begin{equation}
v_{\mathrm{best}} = \frac{\frac{v_A}{\sigma_A^2} + \frac{v_B}{\sigma_B^2}}
              {\frac{1}{\sigma_A^2}   + \frac{1}{\sigma_B^2}}
       = \frac{v_A \sigma_B^2 + v_B \sigma_A^2}{\sigma_A^2 + \sigma_B^2}.
\end{equation}
After a few simplifications, we find that 
\begin{equation}
\chi^2(v_{\mathrm{best}}) = \frac{(v_B - v_A)^2}{\sigma_A^2 + \sigma_B^2}.
\end{equation}

The reduced chi-square is defined as
\begin{equation}
\chi^2_{\mathrm{reduced}} = \frac{\chi^2}{\nu},
\end{equation}
where $\nu$ is the number of degrees of freedom. In our case, we have two measurements and one fitted parameter ($v$); hence, $\nu = 2 - 1 = 1$, so that $\chi^2_{\mathrm{reduced}} = \chi^2$.

This allows us to define a compatibility criterion in terms of an $n\sigma$ agreement. The two estimates are said to be compatible at the $n\sigma$ level if
\begin{equation}
\chi^2 = \frac{(v_B - v_A)^2}{\sigma_A^2 + \sigma_B^2} \le n^2.
\end{equation}

In particular, we can highlight the following agreement levels: $1\sigma$ agreement ($\chi^2 \le 1$), $2\sigma$ agreement ($\chi^2 \le 4$), and $3\sigma$ agreement ($\chi^2 \le 9$).

In practice, we found that none of our measurements agree at the $1\sigma$ level, i.e. $\chi^2 > 1$ for all pairs. We therefore adopted a slightly looser compatibility criterion, requiring $\chi^2 \le 2$, corresponding to a discrepancy of about $1.4\sigma$.

Such a choice is motivated by several considerations. The assumption of purely Gaussian and independent uncertainties is somewhat oversimplified. Systematic errors, such as calibration uncertainties, model dependencies in asteroseismology, or limb-darkening prescriptions in interferometry, are difficult to quantify accurately and may lead to underestimated error bars. For the asteroseismic measurements, although we attempted to account for the dominant sources of systematic uncertainty, we still regard the quoted uncertainties as lower limits. For the interferometric measurements, we relied on values reported in the literature. In addition, a weak level of correlation may be present between the two estimates in some cases, since the interferometric radius is used as a constraint in certain asteroseismic models. These correlations are not reflected in the formal error bars and further motivate the adoption of a more flexible compatibility criterion.

For these reasons, adopting a moderately relaxed threshold such as $\chi^2 \le 2$ provides a pragmatic compromise, allowing us to identify broadly consistent measurements without being overly restrictive in the presence of unaccounted systematics.

\section{Supplementary data for surface modelling}
\label{app:supp_data_surface_modelling}
\begin{figure}[t!]
\centering
\includegraphics[width=.95\linewidth]{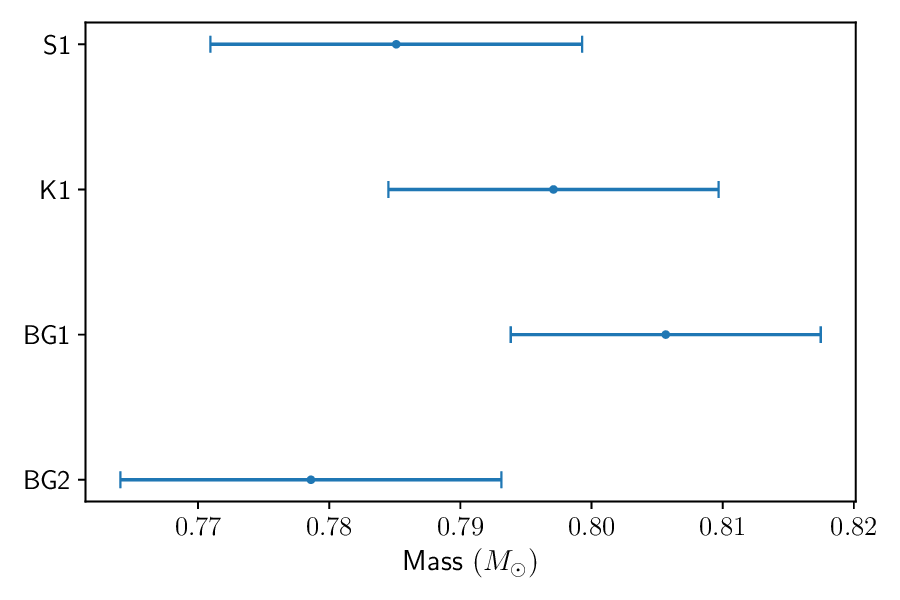}\figsep
\includegraphics[width=.95\linewidth]{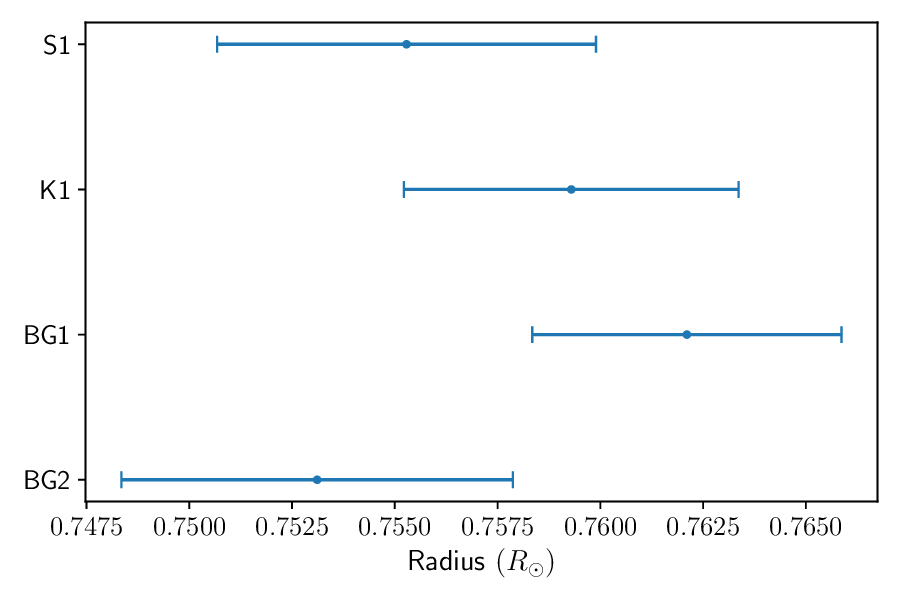}\figsep
\includegraphics[width=.95\linewidth]{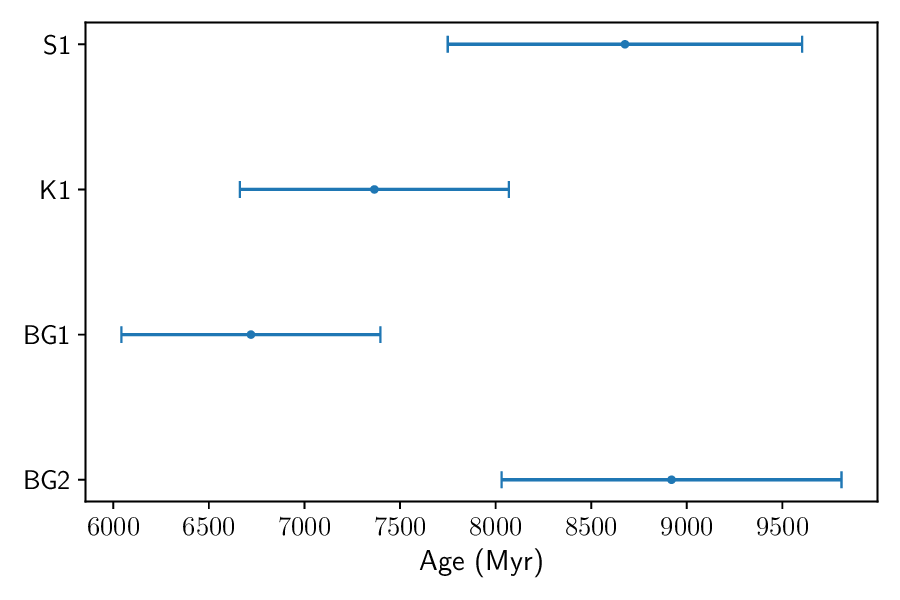}
\caption{Sensitivity of forward-modelling results to different surface effect prescriptions for stellar mass, radius, and age.}
\label{fig:results_forward_and_SE}
\end{figure}

Figure~\ref{fig:results_forward_and_SE} highlights how the choice of the surface effect prescription influences the inferred stellar mass, radius, and age of HD~219134. All major surface effect prescriptions were explored: \citet{Ball&Gizon2014} with one (BG1) or two (BG2) free parameters, \citet{Kjeldsen2008} with one (K1) or two (K2) free parameters, and \citet{Sonoi2015} with one (S1) or two (S2) free parameters. We note that while BG2 and \citet{Sonoi2015} corrections are the most widely used by the community, it is noteworthy that Team 3 in \citetalias{Li2025} employed BG1 prescription for their characterisation of HD~219134.

Minimisations using K2 and S2 prescriptions failed to converge to physical solutions. The remaining prescriptions revealed two distinct behaviours. Substituting S1 for BG2 produces similar results, consistent with the conclusions of \citetalias{Li2025}, though we do not consider these differences negligible relative to the statistical precision achieved for HD~219134 and the PLATO mission accuracy requirements for stellar age. In contrast, BG1 and K1 introduce significant changes, particularly in age, which are much greater than the statistical precision and the PLATO requirements, as discussed in Sect.~\ref{sec:influcence_surface_modelling}.

\section{Transit, orbital, and planetary parameters}
\label{app:transit_orbital_planetary_parameters}

In Table~\ref{tab:planetary_parameters}, we provide the transit, orbital, and planetary parameters of HD~219134b, HD~219134c, HD~219134f, and HD~219134d. Transit and orbital data are sourced from \citetalias{Gillon2017_HD219134}. The planetary parameters were computed as described in Sect.~\ref{sec:revision_planetary_parameters}. 

\begin{table*}[h]
\centering
\caption{Planetary parameters of HD~219134b, HD~219134c, HD~219134f, and HD~219134d.}
\resizebox{0.99\linewidth}{!}{
\begin{tabular}{lccccc}
\hline \hline
Parameter & HD~219134b & HD~219134c & HD~219134f & HD~219134d & Ref. \\ 
\hline 
\textit{Transit and orbital parameters} & & & & & \\ 
Orbital period $P$ (days) & $3.092926\pm 0.000010$ & $6.76458\pm 0.00033$ & $22.717\pm 0.015$ & $46.859\pm 0.028$ & 1 \\
$R_p /R_\star$ & $0.01892\pm 0.00063$ & $0.01783\pm 0.00053$ & >$0.01536\pm 0.00026$ & >$0.01892\pm 0.00018$ &   1 \\
$a_p/R_\star$ & $10.72\pm 0.15$ & $18.05\pm 0.25$ & $40.44\pm 0.56$ & $65.50\pm 0.93$ &  1 \\
Inclination $i$ (deg) & $85.05\pm 0.09$ & $87.28\pm 0.10$ & - & - &  1 \\
Orbital eccentricity $e$ & 0 (fixed) & $0.062\pm 0.039$ & $0.148\pm 0.047$ & $0.138\pm 0.025$ &  1 \\
RV semi-amplitude $K$ (cm/s) & $238.1\pm 7.5$ & $169.7\pm 7.5$ & $192\pm 10$ & $334\pm 11$ &  1 \\
\hline 
\textit{Planetary parameters} & & & & & \\ 
$R_p\, (R_\oplus)$ & $1.554\pm 0.056$ & $1.465\pm 0.048$ & >$1.262\pm 0.028$ & >$1.554\pm 0.027$ &  2 \\ 
$M_p\, (M_\oplus)$ & $4.608\pm 0.199$ & $4.244\pm 0.226$ & >$7.117\pm 0.429$ & >$15.784\pm 0.702$ &  2 \\ 
$\rho_p$ (g/cm$^3$) & $6.745\pm 0.788$ & $7.423\pm 0.833$ & >$19.470\pm 1.743$ & >$23.104\pm 1.567$ &  2 \\ 
$\rho_p\, (\rho_\oplus)$ & $1.22\pm 0.14$ & $1.35\pm 0.15$ & >$3.53\pm 0.32$ & >$4.19\pm 0.28$ &  2 \\ 
$\log g_p$ (dex) & $3.272\pm 0.037$ & $3.288\pm 0.037$ & >$3.642\pm 0.032$ & >$3.807\pm 0.024$ &  2 \\ 
$a_p$ (au) & $0.03754\pm 0.00075$ & $0.06322\pm 0.00125$ & $0.14163\pm 0.00280$ & $0.22940\pm 0.00460$ &  2 \\ 
\hline
\end{tabular}}
{\par\small\justify\textbf{Notes.} The reported uncertainties of the planetary parameters account for the systematics uncertainties of the stellar parameters, which are provided in Sect.~\ref{sec:compatibility_radii}. For HD~219134f and HD~219134d, no measurement of the orbital inclination is available; therefore, only the minimum mass, $M_p \sin i$, can be derived from RV observations. In the table, the reported planetary mass corresponds to the minimum mass (i.e. assuming $i=90$°). \citet{Gillon2017_HD219134} also computed lower limits of $R_p/R_\star$ for planets f and d assuming composition model of \citet{Zeng&Sasselov2013} for a pure iron composition. Consequently, the derived quantities that depend on planetary mass and radius (i.e. mean density and surface gravity) should be regarded as lower limits as well.\\
\textbf{References.} (1) \citet{Gillon2017_HD219134}; (2) this work. \par}
\label{tab:planetary_parameters}
\end{table*}

\section{Position of the Alfven surface}
\label{app:position_Alfen_surface}

Another quantity we derive in our stellar wind model is the position of the Alfven surface, which is relevant for studying magnetic star-planet interactions and their potential associated emissions \citep{Vidotto2025}. Planets orbiting within the Alfven surface may develop a magnetic connectivity with the star. Only our model with the lowest Alfven wave flux allow for planet b to be inside the Alfven surface. In particular, in our preferred model (with intermediate Alfven wave flux) and the model with the highest wave flux, we find that planet b (and thus all other planets beyond that orbit) are in the super-Alfvenic region of the stellar wind. In this case, we do not expect magnetic star-planet interactions to take place in the system, at least, not during the epoch where the magnetic field map was reconstructed. Stellar magnetic fields evolve \citep[e.g.][]{Bellotti2025}, which implies that the stellar wind and its Alfven surface also evolve. There are planetary systems, for example HD 179949, where we expect that planet's b orbit can switch from being super- or sub-Alfvenic as the stellar magnetic field and wind evolve \citep[see e.g. Fig.~\ref{Fig:HD219134_evaporation} and the associated discussion in][]{Vidotto2025}. To further quantify this in the context of HD 219134, additional spectropolarimetric observations and reconstructions of stellar magnetic field at different epochs will be required, in particular in view of the stellar activity cycle of about 13~yr. The ZDI map of HD 219134 is from June-August 2016, which according to Fig. \ref{Fig:S-index_Lx}, is half-way from the rise in activity from its minimum to its maximum. If the stellar magnetic field were to increase towards maximum, then we would expect the Alfven surface to also increase, which would facilitate magnetic star-planet interactions.

\section{Calibration of the braking law from open clusters}
\label{app:calibration_braking_law}
As discussed in Sect.~\ref{sec:rotational_history}, there is a significant difference in spectral type between HD~219134 and the Sun. Therefore, we used a dedicated calibration of the braking law (for both unsaturated and saturated regimes) based on observational data from open clusters with ages between 35~Myr and 4.2~Gyr. Specifically, we sourced data from the following clusters: 
NGC~2547 (35~Myr, \citealt{GodoyRivera2021}), 
Pleiades (120~Myr, \citealt{Rebull2016}), 
M~50 (150~Myr, \citealt{GodoyRivera2021}), 
M~37 (500~Myr, \citealt{GodoyRivera2021}), 
Praesepe (670~Myr, \citealt{Douglas2019}), 
NGC~6811 (950~Myr, \citealt{GodoyRivera2021}), 
NGC~6819 (2.5~Gyr, \citealt{Meibom2015}),
Ruprecht~147 (2.7~Gyr, \citealt{Curtis2020}), and 
M~67 (4.2~Gyr, \citealt{Barnes2016}). We estimated the mass of the cluster members by interpolating absolute magnitudes and dereddened colours in isochrones computed with the Stellar Parameters INferred Systematically software \citep[SPInS;][]{Lebreton2020}, on the basis of a grid of stellar models computed with CLES. For the computation of the grid, we fixed the initial helium mass fraction to $Y_0 = 0.27$, and considered the initial mass ($M_0$) and hydrogen mass fraction ($X_0$) as free parameters, with $0.45 \leq M_0(M_{\odot}) \leq 1.1$, and $0.68 \leq X_0 \leq 0.73$. To derive bolometric corrections in multiple passbands, we relied on the Mesa Isochrones and Stellar Tracks \citep[MIST;][]{Choi2016} project. We selected cluster members with mass in the relatively broad range $0.45 \leq M_{\star}(M_{\odot}) \leq 0.85$. For each cluster, we computed the 25th, 50th and 90th percentile and relative uncertainties of the distribution of surface rotation rates by means of a bootstrap method. For the calibration of the braking law, we varied the initial surface rotation rate ($\Omega_{\rm in}$), the disk-locking timescale ($\tau_{\rm disk}$) and the exponent $\rm p$ considering three types of rotators (fast, moderate and slow), in the attempt to find a reasonable compromise between theoretical rotation tracks and reproducibility of the observational data for the 90th, 50th and 25th percentiles of surface rotation rates distributions, respectively. The best values we found are reported in Table~\ref{tab:braking_law_parameters}.

\begin{table}[h]
\centering
\caption{Braking law parameters calibrated on $0.45 \leq M_{\star}(M_{\odot}) \leq 8.45$ stars.}
\begin{tabular}{lccc}
\hline \hline
Parameter & Fast & Moderate & Slow \\ 
\hline 
$\Omega_{\mathrm{in}}(\Omega_{\odot})$ & 18  & 3.0 & 1.5 \\
$\tau_{\mathrm{disk}}$ (Myr)           & 3.0 & 4.5 & 4.0 \\
p                                      & 1.8 & 1.8 & 1.95\\
\hline
\end{tabular}
\label{tab:braking_law_parameters}
\end{table}

\section{Chromospheric activity and X-ray observations of HD~219134}
\label{app:magnetic_cycle}

\begin{figure}[h]
    \centering
\includegraphics[width=\linewidth]{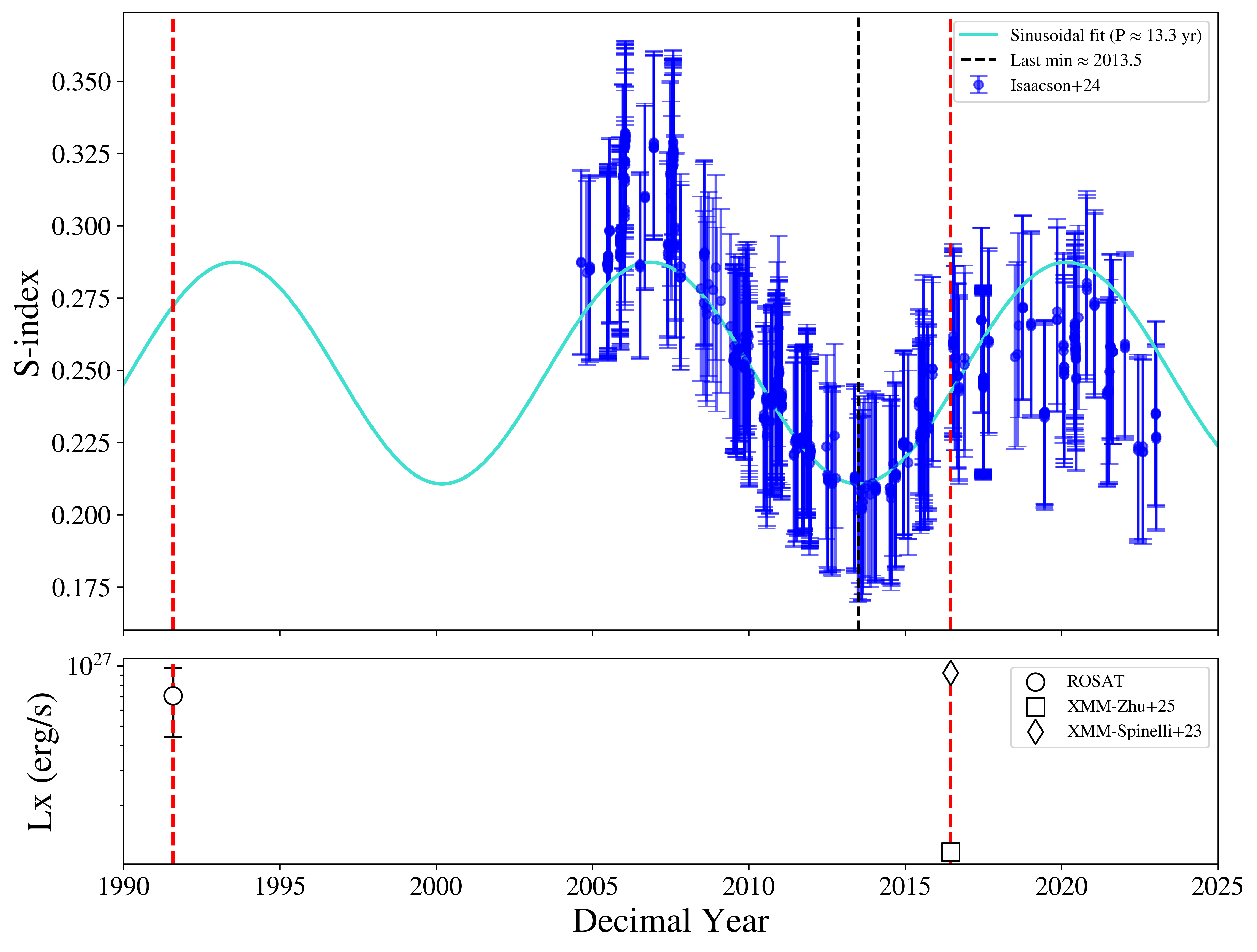} 
    \caption{\textit{Top panel:} Chromospheric activity time-series of HD~219134 from \citet{Isaacson2024} (blue error bars). The light-blue curve shows the best-fit sinusoid, and the dashed black line marks the most recent activity minimum. The vertical dotted red lines indicate the epochs of X-rays observations with ROSAT ($\approx 1991$) and XMM-Newton ($\approx 2016$). \textit{Bottom panel}: X-ray luminosities measured with ROSAT (circle) and XMM-Newton \citep[diamond and square, converted to the ROSAT energy band;][]{Spinelli2023,Zhu2025}.}
    \label{Fig:S-index_Lx}
\end{figure}

In Fig.~\ref{Fig:S-index_Lx} we show a fit of the S-index for HD~219134 from \cite{Isaacson2024} and the relative epochs of observations of the star in the X-ray band by ROSAT and XMM-Newton.

\section{Stellar activity-age relationship}
\label{app:X_age_relationship}

\begin{figure}[h]
    \centering
\includegraphics[width=\linewidth]{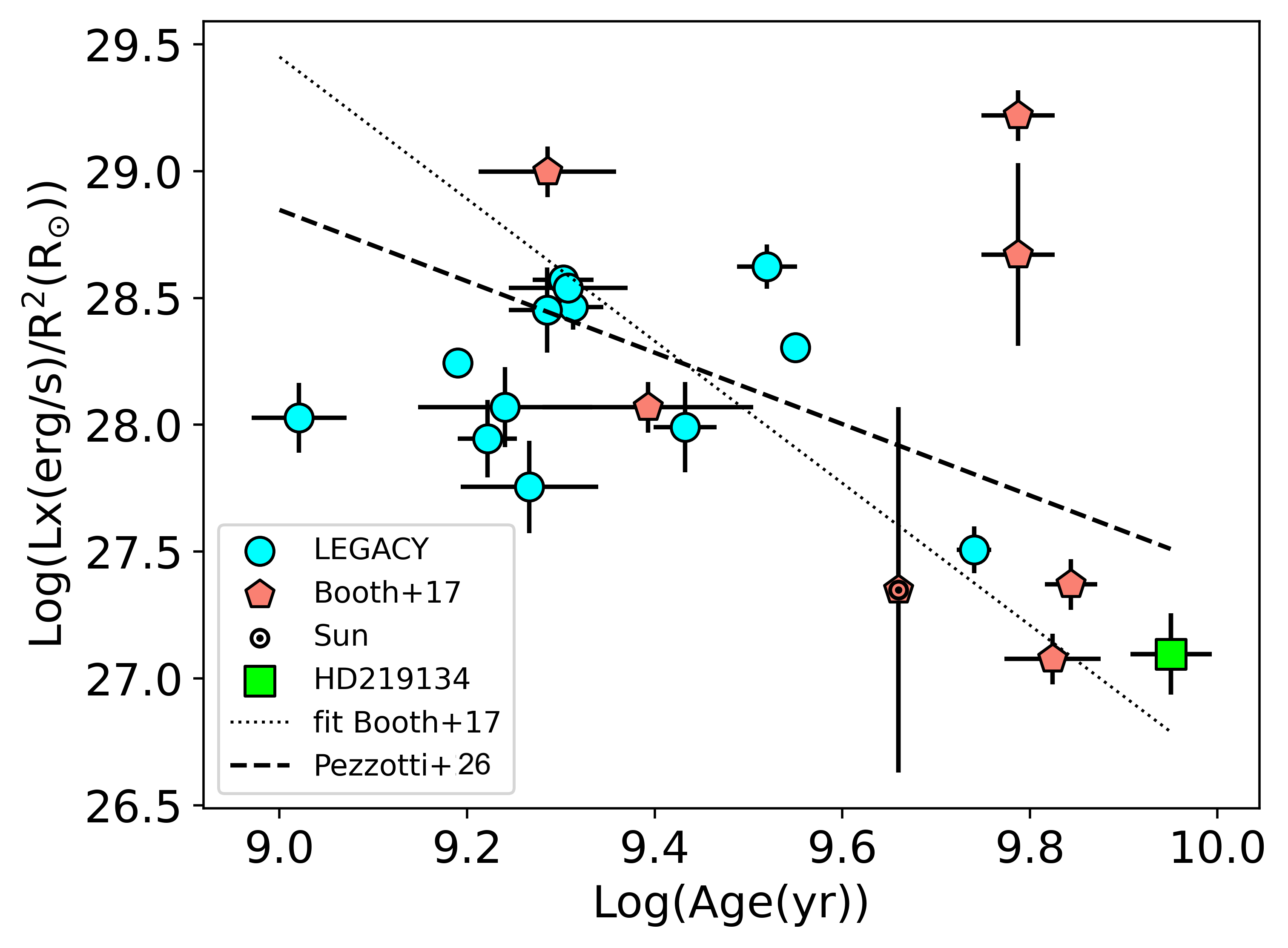} 
    \caption{Activity-age relationship for solar-type stars from \citet{Pezzotti2026}. HD~219134 is denoted by the green square. \textit{Kepler} LEGACY data from \citet{Pezzotti2026} are shown as cyan dots. \citet{Booth2017} data are indicated by salmon pentagons, and the Sun is shown by a circled black dot. The black lines indicate the linear fit from \citet{Booth2017} (dashed line) and \citet{Pezzotti2026} (dotted line).}
    \label{Fig:Activity_Age}
\end{figure}

In Fig.~\ref{Fig:Activity_Age} we indicate the position of HD~219134 in the $\mathrm{Log(L_{x}/R^2)-Log(Age)}$ plane, with respect to other solar-like stars whose ages have been accurately characterised by asteroseismic techniques \citep{Pezzotti2026}.

\end{document}